\documentclass[letterpaper,twocolumn,10pt]{article}
\usepackage{usenix2019_v3}

\usepackage{graphicx}
\graphicspath{{./figures/}}
\DeclareGraphicsExtensions{.pdf, .eps, .mps, .png, .jpg}
\usepackage[english]{babel}
\usepackage{enumitem}
\usepackage{algorithm}
\usepackage[noend]{algpseudocode}
\usepackage[small,compact]{titlesec}
\usepackage{xspace}
\usepackage{xcolor}
\usepackage{amsmath}

\def\HiLi{\leavevmode\rlap{\hbox to \linewidth{\color{black!15}\leaders\hrule height .8\baselineskip depth .5ex\hfill}}}

\makeatletter
\newcommand{\algcolor}[2]{%
  \hskip-\ALG@thistlm\colorbox{#1}{\parbox{\dimexpr\linewidth-2\fboxsep}{\hskip\ALG@thistlm\hspace*{-3pt}\vspace*{-3pt}#2}}%
}
\newcommand{\algemph}[1]{\algcolor{blue!15}{#1}}
\makeatother

\newcommand{\eg}{{\it e.g.,}\xspace}
\newcommand{\ie}{{\it i.e.,}\xspace}

\newcommand{\name}{{ReXCam}\xspace}

\newcounter{packednmbr}

\newcommand{\myparashort}[1]{\vspace{0.05cm}\noindent{\bf {#1}}~}

\begin{document}




\twocolumn[
\bigskip
\centerline{\LARGE \bf ReXCam: Resource-Efficient Cross-Camera Video Analytics at Scale}
\smallskip
\bigskip
\centerline{\large Samvit Jain, Xun Zhang, Yuhao Zhou,}
\smallskip
\centerline{\large Ganesh Ananthanarayanan, Junchen Jiang, Yuanchao Shu, Joseph E. Gonzalez}
\bigskip
\centerline{\normalsize University of California Berkeley, University of Chicago, Microsoft Research}
\bigskip\bigskip
]



\begin{abstract}
Enterprises are increasingly deploying large camera networks for video analytics. Many target applications entail a common problem template: searching for and tracking an object or activity of interest (e.g. a speeding vehicle, a break-in) through a large camera network in live video. Such cross-camera analytics is compute and data intensive, with cost growing with the number of cameras and time. To address this cost challenge, we present ReXCam, a new system for \textit{efficient cross-camera video analytics}. ReXCam exploits spatial and temporal locality in the dynamics of real camera networks to guide its inference-time search for a query identity. In an offline profiling phase, ReXCam builds a \textit{cross-camera correlation model} that encodes the locality observed in historical traffic patterns. At inference time, ReXCam applies this model to filter frames that are not spatially and temporally correlated with the query identity's current position. In the cases of occasional missed detections, ReXCam performs a \textit{fast-replay search} on recently filtered video frames, enabling gracefully recovery. Together, these techniques allow ReXCam to reduce compute workload by 8.3$\times$ on an 8-camera dataset, and by 23$\times$ -- 38$\times$ on a simulated 130-camera dataset. ReXCam has been implemented and deployed on a testbed of 5 AWS DeepLens cameras.

\end{abstract}



\section{Introduction}

The Internet of Things (IoT) has led to an explosion of data sources, and applications that rely on real-time inferences over these data. 
In parallel, the models making these inferences have improved in accuracy, even surpassing humans for certain vision tasks, but at increased resource cost.
This work addresses the systems challenges of scaling up IoT applications to enable {\em live video analytics on a fleet of cameras}.

\begin{figure} [t]
	\centering
	\includegraphics[width=0.8\columnwidth]{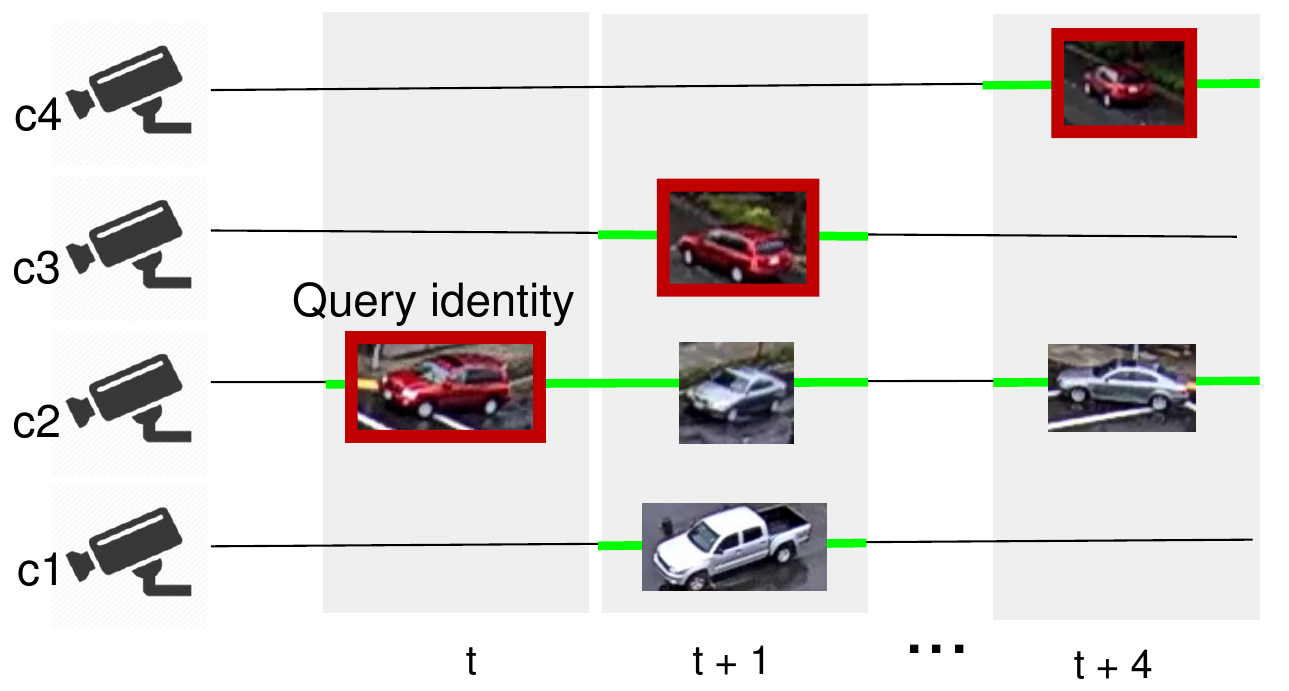}
	\vspace{-0mm}
	\caption{\small \bf \textit{Spatio-temporal correlations} for video inference. The cameras (on the y-axis) are plotted according to their mutual distances, \eg c1 and c2 are spatially closer than c1 and c3. In searching for a query identity starting at frame $t$ (marked in dark red), \name eliminates some cameras entirely (spatial filtering), as well as frames $t+2$ and $t+3$ (temporal filtering). In this example, ReXCam searches first on c1, c2, and c3 (but not c4), finds the target vehicle in c3, and then searches only on c2 and c4 (but not c1 and c3).  The cameras and the times at which they are searched are marked in green. The unmarked portions represent compute savings.}
	\label{fig:example}
	\vspace{-2mm}
\end{figure}

Live video analytics over a fleet of camera feeds embodies two key trends---{\em massive data sources} and {\em compute-intensive inference} (\eg neural nets).
On the one hand, enterprises deploy large camera networks for public safety and business intelligence~\cite{McKinsey}.
For instance, Chicago and London police access footage from 30,000 and 12,000 cameras to respond to crimes in real time~\cite{ChicagoCam,LondonCCTV}. 
On the other hand, many applications rely crucially on cross-camera video analytics, \ie detecting, associating and tracking queried ``identities'' in the live streams as these identities move {\em across the camera feeds over time} (\eg high-value shoppers in a store~\cite{Genetec, jain19hotmobile} or suspects in a city~\cite{Optasia, Vigil}). 
However, cross-camera analytics applications are computationally more challenging than ``stateless'' single-camera vision tasks (such as object detection in one camera feed) as they entail discovering \textit{associations across frames} and {\em across cameras}. Their compute cost thus grows \textit{with} the number of cameras. 


Prior work falls short of addressing this challenge. 
Work in computer vision improves accuracy of cross-camera analytics (\eg~\cite{zhu2018, wang2013intelligent,DukeMTMC}), but it has largely ignored the prohibitive compute costs. 
Recent systems have accelerated analytics on live videos via frame sampling and/or cascaded filters for discarding frames~\cite{VideoStorm, awstream, Chameleon, NoScope,MCDNN,Focus}.
However, they share a key drawback that they optimize the execution of analytics on single video feeds, {\em independent} of the other streams. Thus, the compute cost of cross-camera analytics still grows with more deployed cameras and longer activity time.
\noindent{\bf Spatio-temporal correlations:} Our main insight is that the cost of cross-camera analytics can be drastically reduced by exploiting the {\em physical correlations} of objects among the camera streams. We develop \name, a cross-camera analytics system that leverages inherent {\em spatio-temporal correlations} to aggressively prune the set of camera streams to be processed, thus decreasing compute costs. In the ideal case, \name reduces cost to the number of cameras that the queried object appears in at any point in time and {\em not} the total number of deployed cameras.
A key property of cross-camera applications is that objects of interest appear only in a {\em small number of cameras} at any time, even in large camera deployments. 

Spatial correlations indicate \textit{geographical association} between cameras -- the probability that objects seen in a source camera will move next to a particular destination camera's field of view. Temporal correlations indicate association between cameras \textit{over time} -- the probability that objects seen in a source camera will move next to a destination camera's view \textit{at a particular time}. These {\em spatio-temporal} correlations enable \name to guide its cross-camera inference search toward cameras and frames most likely to contain the {\em query identity} 
(see Figure~\ref{fig:example}). 
\name's use of spatio-temporal correlations to cut the cost of cross-camera analytics is 
fundamentally different than the cross-camera correlations used by recent work (\eg~\cite{Chameleon}) that optimizes the resource-accuracy {\em profiling} but not the live video analytics itself, which still executes on each stream independently. 



\noindent{\bf Challenges:} \name, at its core, applies the physical properties in the IoT world (spatio-temporal correlations across cameras) to high-level AI applications (cross-camera video analytics). 
This has led to three main challenges. 
First, automatically obtaining spatio-temporal correlations is expensive on unlabeled video data.
Second, applying spatio-temporal correlations to existing single-camera inference modules (\eg object trackers) is non-trivial and requires clean abstractions with the necessary system supports. 
Finally, any spatio-temporal profile is bound to have errors that will lead to missing 
objects, which need to be detected and rectified efficiently.

To tackle these challenges, \name operates in three distinct phases. 1) In an offline profiling phase, it constructs a cross-camera \textit{spatio-temporal correlation model} from unlabeled video data, which encodes the locality observed in historical traffic patterns. 
This is an expensive one-time operation that requires detecting entities with an offline tracker, and then converting them into an aggregate profile of cross-camera correlations. 
2) At inference time, \name uses this spatio-temporal model to filter out cameras that are not correlated to the query identity's current position (camera), and is thus unlikely to contain its next instance. 
3) Occasionally, this filtering will cause \name to miss query detections. In these cases, \name performs a \textit{fast-replay search} on recently filtered frames (that it stores), uncovers the missed query instances, and gracefully recovers into its live search.

\noindent{\bf Evaluation Highlights:} We evaluate \name using the well-studied DukeMTMC video data \cite{DukeMTMC} from the Duke campus. On this 8-camera dataset, \name saves compute cost by $8.3\times$ over a correlation-agnostic baseline ($\sim90\%$ of the ideal savings). These savings come at a drop in recall of only $1.6\%$. We also use a simulated dataset of 130 cameras in Porto (using GPS trajectories) \cite{PortoDataset}, and report savings of $23\times-38\times$. Interestingly, \name\ {\em improves} precision by $39\%$, perhaps because the spatio-temporal pruning acts as a ``low pass filter''. Finally, we have implemented and deployed \name on a small testbed of 5 AWS DeepLens smart cameras~\cite{awsdeeplens}. 

\noindent{\bf Contributions:} Our work makes three main contributions. 

\noindent{1) We quantify the potential for harnessing spatio-temporal correlations in cross-camera video analytics.}

\noindent{2) We build a cross-camera video analytics system that learns and applies spatio-temporal profiles on live videos.}

\noindent{3) We develop robust error-handling mechanisms to avoid missed detections by storing and searching on recent videos.}



\section{Motivation and Background}
\label{sec:background}

We explain some example cross-camera video analytics applications (\S\ref{subsec:applications}), the modules in their analytics pipelines (\S\ref{subsec:pipeline}), and then the compute models for video analytics (\S\ref{subsec:setup}).

\subsection{Cross-camera analytics applications}
\label{subsec:applications}

Large camera networks are installed in cities (such as London, Beijing, and Chicago), transport facilities (traffic intersections, airports), and enterprise campuses (corporate offices, retail shops) \cite{UKCam, BeijingCam, ChicagoCam, Vigil}. 
A common class of applications in these camera deployments rely on {\em re-identifying and following objects (\eg people or vehicles) as they move across the views of the different cameras}. The focus is on following select ``objects of interest'' that are typically provided by external entities (such as law enforcement). 
A key characteristic of cross-camera applications is that objects of interest occur only in a {\em small fraction of the cameras} at any given time.

\noindent{\bf 1) Public safety.} Cross-camera video analytics helps localize suspects after a security breach. For example, after a reported incident of a person pulling out a gun inside an office building, 
we will want to track that person (whose image can be obtained from the camera footage) across the cameras in the building while security personnel are dispatched.

Alternatively, after a major public attack (\eg in a train), law enforcement may track the accomplices of the identified perpetrator, which may be obtained from police databases that store people frequently associated with the perpetrator \cite{Vigil}. 
Following these accomplices across the thousands of cameras in the city allows for effective police apprehension.

\noindent{\bf 2) Vehicle tracking in traffic cameras.} In the U.S. and Europe, AMBER alerts are raised on suspected child abductions \cite{AMBER}. The license plate and vehicle details are obtained from investigations, and alerts are broadcast to citizens in the area \cite{AMBER}. Tracking of the suspect's vehicle across the thousands of cameras on highways and city streets can keep tabs on the suspect and victim, even as police intervene \cite{Optasia}.

Likewise, when traffic police notice a vehicle speeding or making a dangerous maneuver, they will note its details and will be interested in tracking the vehicle as it moves across the city using cross-camera analytics to assess its behavior.

\noindent{\bf 3) Retail store cameras.} Using computer vision to improve shopping experience is a big thrust among retailers. ``Special'' shoppers (\eg loyal customers, or customers on wheelchairs) are identified {\em as they enter the store} and cross-camera analytics can be used to track them across the hundreds of cameras in the store to make sure they are provided timely attention (e.g., dispatching a store representative) when necessary. 


\subsection{Video analytics pipelines}
\label{subsec:pipeline}

\begin{figure} [t]
	\includegraphics[width=0.48\textwidth]{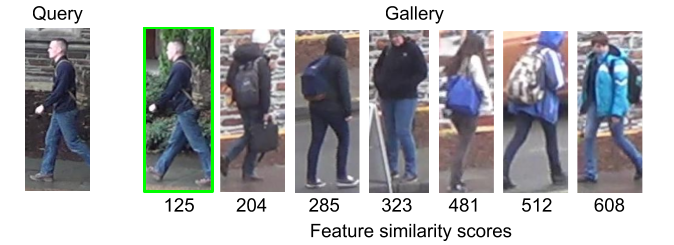}
	\vspace{-6mm}
	\caption{\small \bf Illustration of identity re-identification.}
	\label{fig:re-id}
	\vspace{-4mm}
\end{figure}

Video analytics pipelines for cross-camera applications (in \S\ref{subsec:applications}) typically consist of a series of \textit{modules} on the decoded frames of the video stream: 
(1) an \textit{object detection} module, which extracts and classifies objects of interest in each video frame (\eg people, gun), and (2) a \textit{re-identification} module, which given a query image (\eg of a person), returns positions of co-identical instances of the query in subsequent frames (if present). 
Cross-camera analytics pipelines detect objects in each camera, and track the objects across cameras. Core to this pipeline is the vision primitive of \textit{identity re-identification} \cite{8014818,Ristani18cvpr,5981771}. 
Given an image of a query identity $q$, a re-identification (re-id) algorithm ranks every image in a gallery $G$ based on its \textit{feature distance} to $q$;
the lower the distance the higher the similarity (Figure \ref{fig:re-id}). Typically, features are the intermediate representation of a neural network trained to associate instances of co-identical entities.


Object detection and re-id are the most challenging steps of cross-camera video analytics -- in terms of cost and accuracy -- and our work focuses on improving both of them.


\noindent{\bf Cost.} Tracking in large camera networks is computationally expensive. Tracking even a single object of interest through a camera network, after an initial detection, can potentially require analyzing every subsequent frame in every camera (without good heuristics for geographic localization). \footnote{Optimizations using frame sampling in each camera stream \cite{Focus, NoScope} are orthogonal to our idea of using spatio-temporal correlations across cameras, and we will quantify this aspect in our experiments in \S\ref{sec:results}.}

\noindent{\bf Accuracy.} Re-id is a non-trivial problem in computer vision~\cite{ReidSurvey, MSMT17}, being particularly difficult in crowded scenes and in large camera networks due to significant differences in lighting and viewpoint across cameras. 
Often, re-id models must rely on weak signals (like clothing), thus making it difficult among a large gallery of objects in a frame.

Our use of spatio-temporal correlations to {\em prune} the video frames to analyze -- i.e., run object detection and re-id -- significantly cuts down the inference space, thus improving {\em both cost as well as accuracy}. 
While our focus is on cross-camera applications, we also show how spatio-temporal correlations improve the cost of even single-camera applications (\S\ref{subsec:detection}). 

\subsection{Setup and compute model}
\label{subsec:setup}

Consistent with existing deployments \cite{ieee-computer, videoedge, cloudlets}, our focus is on ``edge'' computation of video analytics. In our setup, all the cameras are in a high-speed {\em local} network with sufficient bandwidth to an edge compute box (e.g., Azure Data Box Edge \cite{adbe}) that is managed by the enterprise (that has deployed the cameras). For example, cameras in an office building are analyzed in an edge box located in the same building. Traffic cameras in a city are analyzed in the local traffic command center \cite{vavz}. Videos are streamed to this edge box and the pipeline modules (\S\ref{subsec:pipeline}) including object detection and re-id are run on this edge. Reducing the compute load enables more video feeds to be processed on the edge box or alternately reduces the resources to be provisioned. 


Our ideas also readily apply to a network of AI cameras (as we implement and deploy in \S\ref{sec:implementation}), each of which consist of compute on-board, accelerators (\eg GPUs), and storage \cite{awsdeeplens, qualcommvip}. 
Our techniques will enable each camera to be provisioned with much lower resources, thus lowering their cost.

\section{Quantifying spatio-temporal correlations}
\label{sec:potential}

We analyze the potential of using spatio-temporal correlations for cross-camera video analytics using the DukeMTMC dataset \cite{DukeMTMC}. We study \textit{cross-camera identity tracking} that involves tracking an object of interest, in real time, through a camera network. In particular, given an instance of a query identity $q$ (\eg a person) flagged in camera $c_q$ at frame $f$, we return all subsequent frames, across all cameras, in which $q$ appears as it moves around. 
We measure the reduction in compute, \ie the number of frames on which object detection and re-id operations (\S\ref{subsec:pipeline}) are executed. 

\subsection{Empirical analysis on cross-camera correlations}
\label{subsec:empirical}

\begin{figure}[t!]
	\centering
	\includegraphics[width=0.45\textwidth]{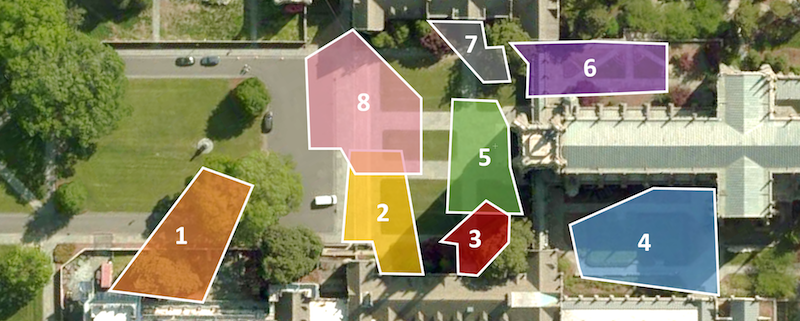}
	\vspace{-1mm}
	\caption{\small \bf DukeMTMC camera network \protect\cite{DukeMTMC}. Marked regions show the visual field of view of each camera.}
	\label{fig:duke-map}
	\vspace{-0mm}
\end{figure}

We now present an empirical study to quantify the \textit{cross-camera correlations} in the DukeMTMC dataset \cite{DukeMTMC}, one of the most popular benchmarks in computer vision person re-id and tracking \cite{Megvii, PersonSearch}. This quantification motivates our design of a video analytics system that leverages such correlations to improve the performance of cross-camera analytics. 
The DukeMTMC dataset contains footage from eight cameras placed on the Duke University campus (see Figure \ref{fig:duke-map}), in an area with significant pedestrian traffic. The field of views of the cameras do not mostly intersect, but the cameras are placed close enough that people frequently appear in multiple cameras, as is typical in camera deployments. The dataset contains over 2,700 unique identities across 85 minutes of footage, recorded at 60 frames per second \cite{DukeMTMC}.

\subsubsection{Spatial correlation.} \label{subsubsec:spatial}
\begin{figure}[t!]
	\includegraphics[width=0.485\textwidth]{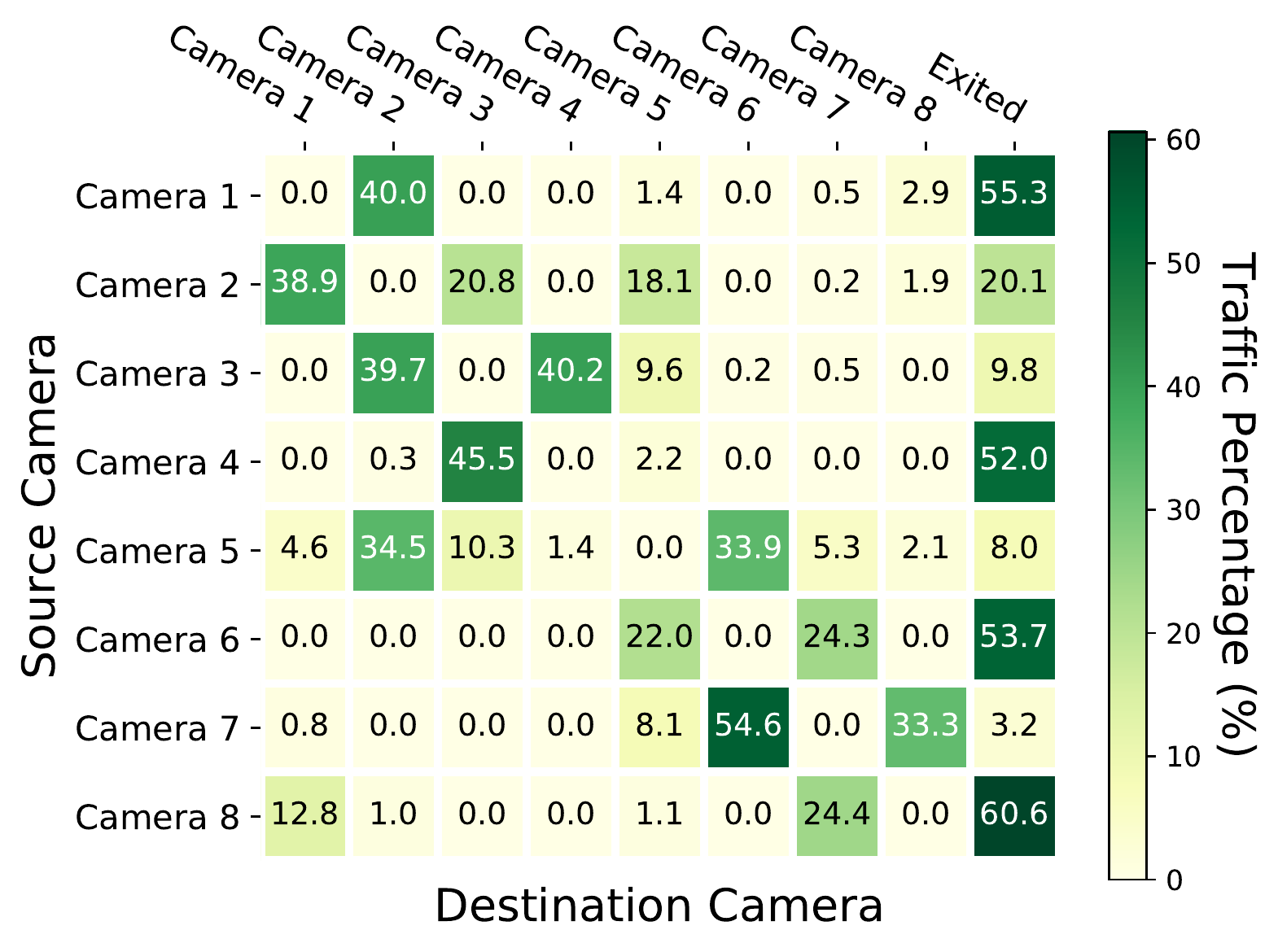}
	\vspace{-8mm}
	\caption{\small \bf Spatial correlations in the DukeMTMC dataset \protect\cite{DukeMTMC}. Cells display \% of outbound traffic (individuals) from each camera that appears at other cameras. Each row corresponds to a particular source camera while each column to a destination camera; each row's values add up to 100\%. The final column represents traffic that \textit{exits} the camera network.}
	\label{fig:spatial}
	\vspace{-4mm}
\end{figure}

Cross-camera movement of individuals (or ``traffic'') demonstrates a high degree of spatial correlation. Here, ``traffic'' between cameras $A$ and $B$ is defined as the set of unique individuals detected in camera $A$ that are {\em next} detected in camera $B$. (Note that a person that moves from $A$ to $B$ via camera $C$ are excluded from the traffic count of $A \rightarrow B$ and instead included in the $A \rightarrow C$ traffic count.) We find that individuals seen at a camera $c_q$ move next to only a small number of $c_q$'s peer cameras. 
On the 8-camera DukeMTMC dataset, only 1.9 of 7 potential peer cameras, on average, receive {\em even 5\%} of the total outbound traffic (or individuals) from a given camera. 
Figure \ref{fig:spatial} shows the full pair-wise spatial correlations. 

Exploiting this insight can significantly reduce our compute workload, at little cost to accuracy, when searching for a query identity $q$ (\eg a person), that was first detected in camera $c_q$. In comparison to a scheme that searches all $n-1$ peers, a smarter scheme that searches only those camera feeds that receive \textit{at least 5\%} of the traffic from $c_q$, reduces our compute by $3.7\times$ (we search only 1.9 cameras instead of 7, or $3.7\times$ fewer frames to run object detection and re-id; see \S\ref{subsec:pipeline}), while {\em still} capturing $95\%$ of all detections as per our experiments. 

An interesting aspect is that geographical proximity is {\em not} necessarily a good spatial filter. Consider camera-5 (Figure \ref{fig:spatial}), out of which a significant fraction of individuals (traffic) go to cameras 2 and 6 {\em but not to} 7 or 8 even though they are also spatially proximate. Likewise, little traffic moves out of camera 8 to cameras 2 and 5 even though these are physically proximate. Thus, learning these patterns in a data-driven fashion is a more robust approach (as we will quantify in \S\ref{sec:results}). Data-driven learning also allows us to capture asymmetry in the traffic patterns between cameras, for e.g., over 50\% of traffic from camera-7 move to camera-6 but less than 25\% of traffic moves in the reverse direction from camera-6 to 7.

\begin{figure}[t!]
	\includegraphics[width=0.485\textwidth]{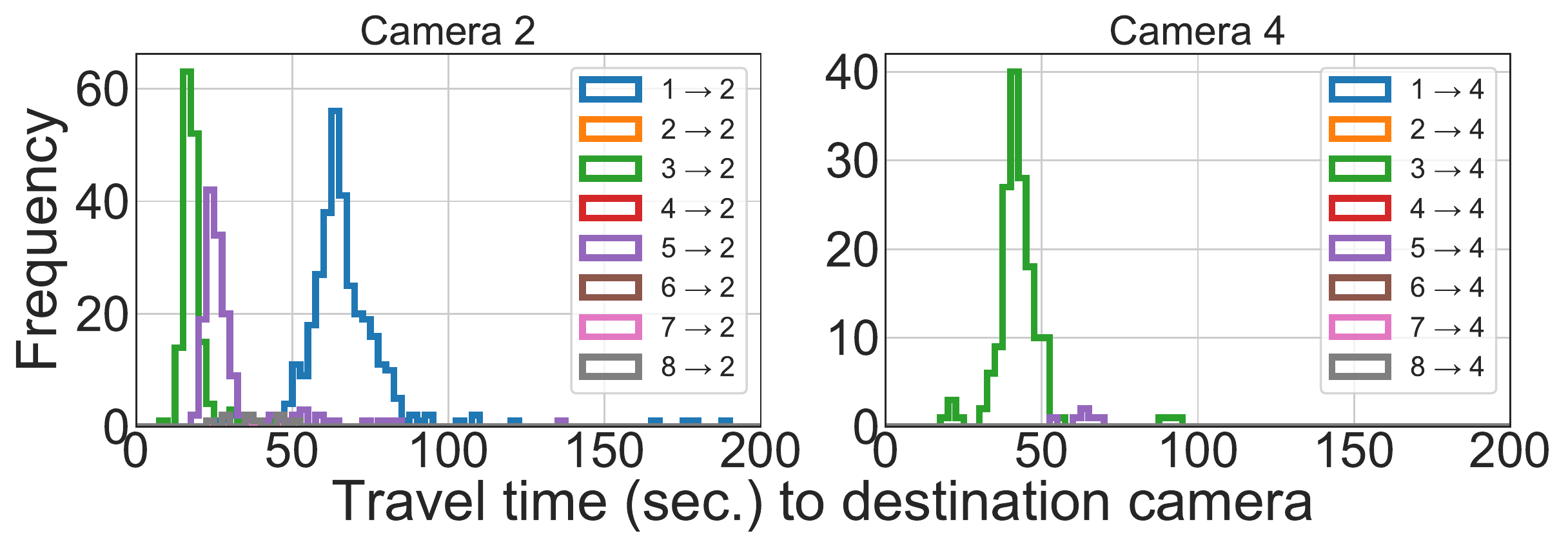}
	\vspace{-8mm}
	\caption{\small \bf Temporal correlations in the DukeMTMC dataset \protect\cite{DukeMTMC} (for two example destination cameras 2 and 4). Plots display distribution of inter-camera travel times. Each plot corresponds to traffic to the particular destination camera. Each colored line represents a particular source camera.}
	\label{fig:temporal}
	\vspace{-4mm}
\end{figure}

\subsubsection{Temporal correlation.}\label{subsubsec:temporal}
Cross-camera traffic also demonstrates a high degree of \textit{temporal} correlation. As Figure \ref{fig:temporal} shows, travel times of individuals between a particular source camera and a destination camera in the DukeMTMC dataset are highly correlated. This is explained by the fact that these are static cameras and thus their pairwise distances are also static. 
Thus, for a given pair of cameras, the travel times for people to leave the feed of one camera and appear in the other camera are likely to be clustered around a mean value. 
In the DukeMTMC dataset, the average travel time between all camera pairs is 44.2s, and the standard deviation is only 10.3s (or only 23\% of the mean).


Exploiting temporal correlations, even on its own, has the potential to provide compute savings. Given the task of locating a given query identity $q$, first identified in camera $c_q$, in one of the $n-1$ possible destination camera streams, we can simply search {\em each of the} $n-1$ streams (ignoring spatial correlations) but only for the time window when the query identities are most likely to show up. We probabilistically set the time window to be when {\em at least 98\%} of the objects appear. Such an approach has the potential to reduce our compute load by $7.5\times$ compared to a naive approach that does not use such a (time) windowed search. This shows the considerable potential in leveraging the tight distribution of travel times of individuals between the views of the cameras.

\subsection{\bf Potential gains: spatial \& temporal correlations}

We now put together the gains due to spatial and temporal filtering combined over a baseline that searches all $n-1$ cameras (for a maximum duration). 
We assume ideal knowledge about the spatial correlations between the cameras as well as the temporal characteristics of travel times of individuals between the views of the cameras. Using the same thresholds as in \S\ref{subsec:empirical}, our analysis shows a potential gain of $9.4\times$ savings in the compute cost. This encouraging potential for savings, even for a 8-camera dataset, motivates us to both learn and exploit the spatio-temporal correlations for cross-camera video analytics. 
As we will show in \S\ref{sec:evaluation}, \name achieves $8.3\times$ reduction in compute cost, which is $\sim90\%$ of the potential. In addition, the filtering of frames to search also improves the {\em precision} of the results from 51\% for the baseline approach to 90\% with \name, with little drop in recall.

\section{\name Overview}

\begin{figure}[t!]
	\centering
	\includegraphics[width=0.4\textwidth]{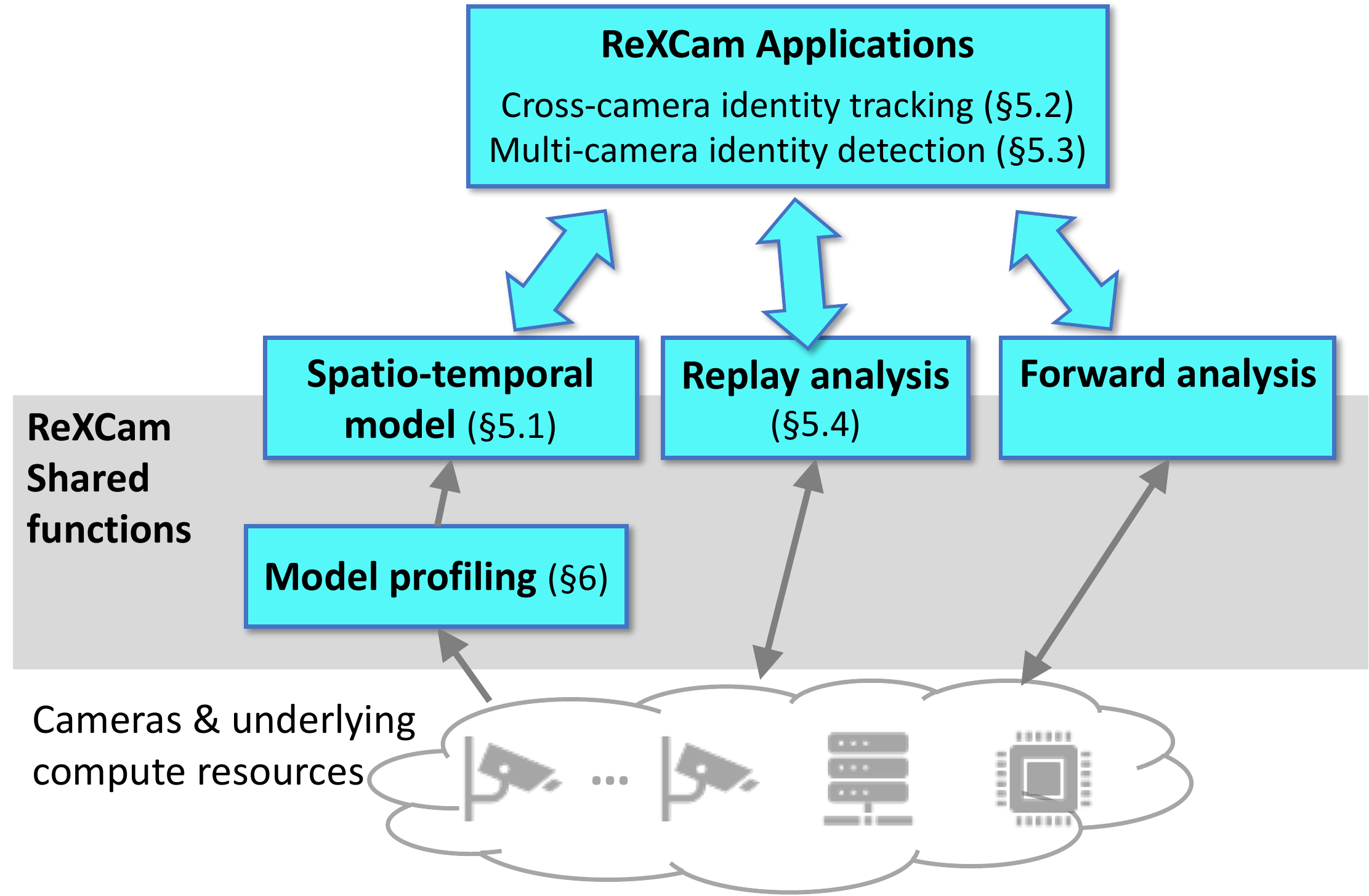}
	\vspace{-2mm}
	\caption{\small \bf Architecture of \name.}
	\label{fig:overview}
	\vspace{-4mm}
\end{figure}

Building upon the strong spatial/temporal correlations across cameras seen in \S\ref{sec:potential}, we develop \name, a {\bf r}esource-{\bf e}fficient {\bf cross}-{\bf cam}era analytics system that leverages the correlations across cameras to reduce computing cost.
As depicted in Figure~\ref{fig:overview}, \name provides two core functions for cross-camera video analytics applications. 

\vspace{0.1cm}
\noindent {\bf The spatio-temporal model} (\S\ref{subsec:st-model}) describes the spatial and temporal correlation between cameras, and can be queried by applications. 
At a high level, one can query the model with two cameras, $c_s$ and $c_d$, and a time window, and it will return how likely an object leaving $c_s$ will appear in $c_d$ (\ie the spatial correlation) and if it appears in $c_d$ how likely it will appear within the time window (\ie temporal correlation).

\vspace{0.1cm}
\noindent {\bf The forward and replay analysis} (\S\ref{subsec:inference} and \S\ref{sec:replay}) perform real-time inference on live videos (\ie forward) as well as inference on history video (\ie replay). 
Both capabilities operate jointly, and replay search is inherently needed for spatio-temporal pruning: ignoring a camera due to weak spatial/temporal correlation will inevitably introduce false negatives that a baseline of searching all cameras would have avoided, so \name provides the abstraction of replay search to allow faster-than-real time search over some history videos (that were ignored) for error correction.


In \S\ref{subsec:inference} we demonstrate how cross-camera identity tracking (tracking an identity across cameras over time from a known starting point) is performed using spatio-temporal pruning. We also show the generality of the functionalities of \name by applying spatio-temporal pruning for cross-camera identity {\em detection} (finding a queried identity, e.g., a lost child, in a large camera deployment) in \S\ref{subsec:detection} that is both an important {\em single-camera} application as well as ties to the cross-camera identity tracking by providing it the starting point for its tracking.

\section{Spatio-temporal correlations in \name}
\label{sec:st}


We now describe \name's solution for leveraging spatio-temporal correlations in cross-camera video analytics.

\subsection{Defining the spatio-temporal model}
\label{subsec:st-model}
\name builds upon the cross-camera correlations in \S\ref{sec:potential}.

\noindent{\bf 1) Spatial correlations} capture associations between camera pairs arising from the movement of traffic (individuals) between the views of the camera streams. The degree of spatial correlation $S$ between two cameras $c_s, c_d$ is quantified by the ratio of: (a) the number of individuals leaving the source camera's stream for the destination camera, $n(c_s, c_d)$, to (b) the total number of entities leaving the source camera:
\vspace{-0.07in}\begin{align*}
	S(c_s, c_d) = \frac{n(c_s, c_d)}{\sum_i n(c_s, c_i)}
\end{align*}
When a large fraction of individuals that leave $c_s$'s view are seen next in a camera $c_i$, we say that $c_i$ is \textit{highly correlated} to camera $c_s$. Note that S may be asymmetric (as seen in our analysis in \S\ref{subsubsec:spatial}); camera $c_s$ may \textit{not} be highly correlated with camera $c_i$, even if the converse is true. In cross-camera identity search, ReXCam exploits spatial correlations by prioritizing cameras that are highly correlated to the last camera where the queried identity $q$ was spot (called {\em query camera}).

\noindent{\bf 2) Temporal correlations} capture associations between camera pairs \textit{over time}. If a large fraction of the traffic leaving camera $c_s$ for camera $c_d$ arrives within durations $t_1$ and $t_2$, then camera $c_d$ is said to be \textit{highly correlated in the time window $[t_1, t_2]$} to camera $c_s$. The degree of temporal correlation $T$ between two cameras $c_s, c_d$ during a window $[t_1, t_2]$ is the ratio of: (a) individuals reaching $c_d$ from $c_s$ within a duration window $[t_1, t_2]$ to (b) total individuals reaching $c_d$ from $c_s$:
\vspace{-0.07in}\begin{align*}
	T(c_s, c_d, [t_1, t_2]) = \frac{n(c_s, c_d, [t_1, t_2])}{n(c_s, c_d)}
\end{align*}
Indeed, cameras in real-world deployments have substantial temporal correlation (\S\ref{subsubsec:temporal}). 
In cross-camera identity search, ReXCam exploits temporal correlations by prioritizing the \textit{time window} $[t_1, t_2]$ in which a destination camera is most correlated with the query camera.

\noindent{\bf Spatio-temporal model} 
Given a source camera $c_s$, the current frame index $f_{\text{curr}}$ (which serves as a timestamp), and a destination camera $c_d$, our proposed spatio-temporal model $M$ outputs \texttt{true} if $c_d$ is both spatially and temporally correlated with $c_s$ at $f_{\text{curr}}$, and \texttt{false} otherwise. In our description, the frame index $f_{\text{curr}}$ serves the role of the timestamp.

\begin{figure}
	\centering
	\includegraphics[width=1.0\columnwidth]{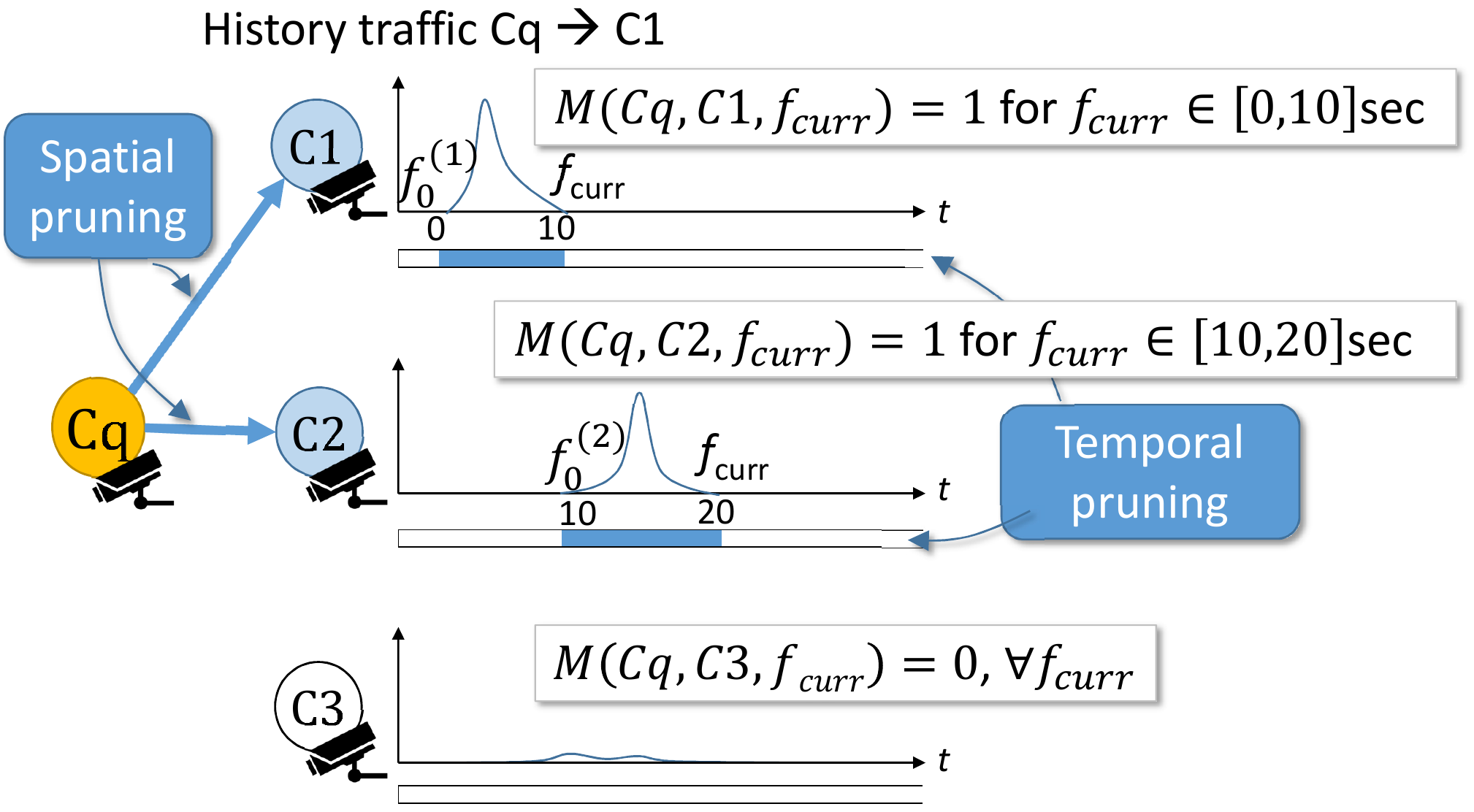}
	\vspace{-5mm}
	\caption{\small \bf
	Spatio-temporal correlations between camera Cq (where the object was first spotted) and three other cameras. 
	C1 and C2 have spatial and temporal correlations with Cq (in different time intervals). C1 is correlated with Cq in the times [0, 10] but not otherwise; and C2 is correlated with Cq only in times [10, 20]. C3 is not correlated with Cq.
	}
	\label{fig:illustration}
	\vspace{-2mm}
\end{figure}

The thresholds for being spatially correlated with $c_s$, and temporally correlated with $c_s$ at time $f_{\text{curr}}$ are model parameters. As an example, we may first wish to search cameras receiving at least $s_{\text{thresh}} = 5\%$ of traffic from $c_s$, during the time window containing the first $1 - t_{\text{thresh}} = 98\%$ of traffic from $c_s$. These parameter settings exclude both \textit{outlier cameras} (cameras receiving less than $5\%$ of the traffic from $c_s$) and \textit{outlier frames} (frames containing the last 2\% of the traffic from $c_s$). 
Defining $s_{\text{thresh}}$ and $t_{\text{thresh}}$ as a percent of traffic (or individuals) directly translates to precision and recall of the entities being tracked. 
$M$ is formally defined as:
\begin{align}
M(c_s, c_d, f_{\text{curr}})& = \label{eq:model}
	\begin{cases}
		1, & S(c_s, c_d) \geq s_{\text{thresh}}\\
		& \quad \textbf{and} \\
		& T(c_s, c_d, [f_0, f_\text{curr}]) \leq 1 - t_{\text{thresh}}\\
		0, & \text{otherwise}
	\end{cases}
\end{align}
Here $f_0$ is the frame index at which the first historical arrival at $c_d$ from $c_s$ was recorded. The reason of having $f_0$ is because it takes time to travel from $c_s$ to $c_d$, and cost savings can be maximized by not searching on frames while objects are moving between cameras. As a result, our temporal filter checks if the volume of historical traffic that arrived at $c_d$ between $[f_0, f_\text{curr}]$ is less than $1 - t_{\text{thresh}}$ of the total traffic. This ensures that $f_\text{curr}$ falls in the ``dense'' part of the travel time distribution, where we are likely to find $q$. (Note that we must check that $f_\text{curr} \geq f_0$. When $f_\text{curr} < f_0$, $M$ is \texttt{false}.) Figure \ref{fig:illustration} shows an illustration for using $M$ with $f_0$ values for each destination camera. (We construct the model $M$ in \S\ref{sec:training}.)

\myparashort{Search hits and misses:} Leveraging the spatio-temporal model $M$ 
allows us to explore the subset of the inference space (camera streams and time windows) that is most likely to contain $q$. A ``hit'' reduces cost, as we avoid searching the entire space. 
On the (rare) misses, we go back and find $q$ in the {\em past} video frames over all the camera streams we had filtered out using $M$. In \S\ref{sec:replay}, we will explain how we handle misses and mitigate the {\em delay} it introduces. Maximizing the cost savings from hits and minimizing the miss-induced delays is a tradeoff controlled by the parameters $s_\text{thresh}$ and $t_\text{thresh}$.

\begin{algorithm} [t!]
	\caption{Tracking with the spatio-temporal model}
	\label{alg:tracking-st}
	\begin{algorithmic}[1]
	    \small
		\State \textbf{input}: video feeds $\{V_c\}$ for camera $c$,
		\\ \algemph{\quad\quad\quad $\text{sp\_corr}(c_s, c_d) \rightarrow \{\texttt{true}, \texttt{false}\} $}
		\\ \algemph{\quad\quad\quad $\text{tp\_corr}(c_s, c_d, f) \rightarrow \{\texttt{true}, \texttt{false}\} $}
		\For{query $(q, f_q, c_q) \in Q$}
		\State $q_{\text{feat}} = \text{features}(q)$ \Comment{extract image features}
		\State $f_{\text{curr}} = f_q + 1$ \Comment{init current frame index}
		\State $M_q = []$ \Comment{init query match array}
		\State \algemph{$\text{phase} = 1$ \Comment{start phase one}}
		\While{$(f_{\text{curr}} - f_q) \leq \text{exit\_t}$}
		\State \algemph{$V_{\text{corr}} = \textbf{filter}(\text{sp\_corr}, \text{tp\_corr}, c_q, f_{\text{curr}}, V)$}
		\State $\text{frames} = \text{get\_frames}(V_{\text{corr}}, f_{\text{curr}})$
		\State $\text{gallery} = \text{extract\_entities}(\text{frames})$
		\State $\text{ranked} = \text{rank\_reid}(q_{\text{feat}}, \text{gallery})$
		\If{ $\text{ranked}[0][\text{dist}] < \text{match\_thresh}$}
		\State $M_q = \text{append}(M_q, \text{ranked}[0][\text{img}])$
		\State $q_{\text{feat}} = \text{update\_rep}(q_{\text{feat}}, \text{ranked}[0][\text{feat}])$
		\State $f_q = f_{\text{curr}}$
		\State \algemph{$\text{phase} = 1$ \Comment{reset to phase one}}
		\State $\text{break}$
		\EndIf
		\State $f_{\text{curr}} = \text{increment}(f_{\text{curr}})$
		\If{$\text{phase} = 1 \textbf{ and } T(c_s, c_d, [f_0, f_\text{curr}]) > 1 - t_{\text{thresh}}$}
		\State \algemph{$f_{\text{curr}} = f_q + 1$ \Comment{reset frame index}}
		\State \algemph{$\text{sp\_corr} = \text{relax}(\text{sp\_corr})$}
		\State \algemph{$\text{tp\_corr} = \text{relax}(\text{tp\_corr})$}
		\State \algemph{$\text{phase} = 2$ \Comment{start phase two}}
		\EndIf
		\EndWhile
		\EndFor
		\State \textbf{output}: matched detections $\{M_q\}$
	\end{algorithmic}
\end{algorithm}

\subsection{
Cross-camera identity tracking}
\label{subsec:inference}
Algorithm \ref{alg:tracking-st} explains our \textit{cross-camera identity tracking}. In cross-camera identity tracking, the input consists of a query image $q$, last seen in frame $f_q$ on camera $c_q$. (If the input does not contain the frame $f_q$, we can first run the next application, multi-camera identity detection, to locate it.)
The goal is to flag all subsequent frames, on all cameras, where $q$ appears. 
Note that $q$ can appear again on the same camera ($c = c_q$), different cameras ($c \neq c_q$), or else exit the network altogether. 
For each query $q$, we begin by extracting image features $q_{\text{feat}}$ and initializing an empty array of discovered matches $M_q$. For each frame, as explained in \S\ref{subsec:pipeline}, we: 
(1) extract individuals (objects) from each frame using an object detection model, 
(2) rank the objects based on their feature similarity distance to $q$ using a re-id model (Figure \ref{fig:re-id}).

If the top-ranked detection is within a threshold ({\em match\_thresh} in Algorithm \ref{alg:tracking-st}), i.e., a co-identical instance is found by the re-id model, we add the detection to our array of matches $M_q$, update our query \textit{representation} $q_{\text{feat}}$ to incorporate the features of the new instance of $q$, update the query frame index $f_q$ to $f_{\text{curr}}$, and proceed with tracking $q$; lines 14-18. 
We continue searching until the gap between the last detected instance of $q$ and our current frame index exceeds a pre-defined \textit{exit threshold} (defined as $\text{exit\_t}$ in Algorithm \ref{alg:tracking-st}). At this point, we conclude that $q$ must have exited the camera network, and cease tracking $q$. 

We apply the spatio-temporal model to cross-camera tracking as follows (marked in blue in Algorithm \ref{alg:tracking-st}). The model $M$ has two filters (lines 2 and 3): 
(1) $\textbf{spatial\_corr}(c_s, c_d)$, which given a source camera $c_s$ and a destination camera $c_d$ returns \texttt{true} if $c_d$ is correlated with $c_s$, and (2) $\textbf{temporal\_corr}(c_s, c_d, f)$, which given a source camera $c_s$, a destination camera $c_d$, and a frame index $f$, returns \texttt{true} if $c_d$ is correlated with $c_s$ at $f$. At query time, these two functions are passed to the \textbf{filter} function (line 10), which given a list of video feeds $V$, returns the subset of cameras ($V_{\text{corr}}$) that are \textit{both} spatially and temporally correlated to $c_q$ at $f_{\text{curr}}$.

Applying \textbf{filter} reduces the inference search space, at each frame step $f_{\text{curr}}$, from all entity detections at $f_{\text{curr}}$ on every camera to all entity detections at $f_{\text{curr}}$ on \textit{correlated} cameras. This allows us to abstain from running object detection and feature extraction models on non-correlated cameras, and reduces the size of the re-id gallery in the ranking step. 
If \textbf{filter} in Algorithm \ref{alg:tracking-st} were applied to the example in Figure \ref{fig:illustration}, the set $V_{\text{corr}}$ would be only C1 in in the times [0, 10], only C2 in the times [10, 20], and null set at all other times.


\subsection{Handling pruning errors via replay search}
\label{sec:replay}


Spatio-temporal pruning may cause a drop in recall: {\em missing actual occurrences} of the query identity $q$, which would be discovered by a baseline that exhaustively searches all the frames of all the cameras. 
When tracking on the spatially filtered cameras does {\em not} discover $q$ after exit\_t time (line 22 in Algorithm \ref{alg:tracking-st}), we will initiate a ``second pass'' through the video frames 
that we skipped; we call this \textit{replay search}.


\noindent{\bf Replay subset:} We initiate replay search on a broader subset of cameras and timespans.
In particular, we go back to the last camera that the queried identity was seen, $c_q$ (\ie restart the tracking procedure from $f_{\text{curr}} = f_q + 1$, line 23, as $f_q$ was the last frame the queried object was seen), and find all the correlated cameras and time windows that $c_q$ is correlated with using the spatio-temporal profile {\em but} now with thresholds s$_\text{thresh}$ and t$_\text{thresh}$ decreased by a factor of 10.
If we do discover an instance of $q$, we proceed with tracking from that detection, initiating a new phase one in Algorithm \ref{alg:tracking-st}. If we still do not, we search the entire camera network until the exit threshold.


Note that despite relaxing s$_\text{thresh}$ and t$_\text{thresh}$, the cameras over which we perform replay search will still be only a small fraction of the overall camera network and for only a small duration in the past. This is because a vast majority of cameras (in a large deployment) will have never seen traffic (individuals) from $c_q$. 
Implicit to replay search is also the ability to store videos in the past. However, this only needs to be for the last few minutes (few 100 MBs even for HD videos).

\noindent{\bf Replay delay:} Searching on videos from the past indicates that we are lagging behind tracking the identity. 
Thus, it is desirable to speed up the search process.
\name processes the {\em historical videos} at {\em faster-than-real-time}. 

{\em a) Skip frame mode} -- Process the historical videos at lower frame rate (via frame sampling) and lower resolution (via frame downsizing) to increase processing rate but potentially lower accuracy. We use offline profiling \cite{awstream, VideoStorm} to decide the frame rates and resolution to limit the drop in accuracy. 

{\em b) Parallelism mode} -- Process the historical videos by parallelizing them across other cameras or edge machines (depending on the setup; \S\ref{subsec:setup}) that are idle. As explained above, the broader replay search is likely still only a small subset of all the videos, so spare resources will be available. 

We implement both solutions and investigate their trade-offs on accuracy and delay in our evaluation (\S\ref{sec:eval-frs}). 


\subsection{
Multi-camera identity detection}\label{subsec:detection}

While our focus thus far has been on cross-camera video analytics, spatio-temporal models can also be applied to reduce the cost of {\em single-camera analytics}, e.g., find a lost baby or lost car in a mall's or city's cameras. This involves running object detectors {\em independently on each camera stream}, and is expensive for large camera deployments. In this section, we apply our cross-camera spatio-temporal model (\S\ref{subsec:st-model}) to such single-camera ``identity detection''. 
Not only is it an application of wide relevance on its own, it also ties closely with cross-camera tracking (\S\ref{subsec:inference}) to provide it the starting point of the query $q$ (which we have been referring to as camera $c_q$). 

Identity detection refers to finding a given identity $q$ (\eg an image of a lost baby or suspect) in many camera streams. 
The intuition why the spatio-temporal model helps is that if $q$ is not found in camera C1 and the spatio-temporal model indicates that most objects appearing in camera C2 have recently appeared in C1, then camera C2 is unlikely to contain $q$. 
In other words, 
the model allows to prune the cameras and time windows in which $q$ is {\em unlikely} to be found based on when and where $q$ was {\em not} found earlier.
At any point of time, we maintain a probability for each camera to contain an object that has not been ``scanned'' (\ie not found in the camera feeds we have searched so far).
The cameras with high values of this probability will be prioritized in the search.

Formally, we define $P_{c,w}$ to be the probability of any unscanned object (\ie an object that did not appear in any camera when it was searched) appearing in camera $c$ in time window $w$. 
Thus, the greater the $P_{c,w}$ is, the more likely searching camera $c$ in window $w$ would yield a ``hit''.
We also define $P^*_c$ is the probability of the identity entering the whole camera network at camera $c$ at any point in time.
We estimate this value by looking at the history trace and dividing the number of objects who appear camera $c$ first by total number of objects. 
Then $P_{c,0}=P^*_c$ and $P_{c,w}$ with $w>0$ can be derived iteratively by the following equation:
\begin{align*}
P_{c,w} = P^*_c+ \sum_{w_j\leq w,c_i}I_{c_i,w_j}\cdot P_{c_i,w_j}\cdot S(c_i,c)\cdot T(c_i,c,w)
\end{align*}
where $I_{c_i,w_j}$ is a binary flag indicating if camera $c_i$ was searched at time window $w_j$ ($I_{c_i,w_j}=0$) or not ($I_{c_i,w_j}=1$).
The equation can be intuitively interpreted as following:
the probability of query object $q$ to appear in camera $c$ and time window $w$ is the sum of the probability of it entering the whole network at $c$ (\ie $P^*_c$) and
the probability of $q$ moving from another camera $c_i$ to at time $w_j$, \ie $I_{c_i,w_j}\cdot P_{c_i,w_j}\cdot S(c_i,c)\cdot T(c_i,c,w)$.

At any point in time, we search the camera $c$ and time window $w$ whose $P_{c,w}$ is greater than a threshold $\theta$.
If the identity is found, the search ends. Otherwise, we set $I_{c_i,w_j}=0$ and update other $P_{c,w}$.
This is run until we find the queried identity. \S\ref{sec:iddet-eval} evaluates our gains with identity detection.

\section{Profiling spatio-temporal correlations}
\label{sec:training}




A final piece of ReXCam system is the profiling and maintaining of the spatio-temporal correlations. 
ReXCam takes an approach that builds on standard techniques from computer vision.
Before ReXCam is deployed, we first use a \textit{multi-target, multi-camera} (MTMC) \textit{tracker} to label entities in a dataset of historical video, collected from the same camera deployment on which the live tracking is executed. 
Logically, such a tracker will return for each detected entity instance $i$ a tuple, $(c_i, f_i, e_i)$, containing the camera identifier $c_i$, frame index $f_i$, and entity identifier $e_i$ for the detection, respectively.

Using these, we compute $n(c_s, c_d, [t_1, t_2])$, the number of entities leaving any source camera $c_s$ for any destination camera $c_d$ within a time interval $[t_1,t_2]$.
These quantities translate directly to our spatio-temporal model $M$ in Eq. \ref{eq:model} (see \S\ref{subsec:st-model}). 

However, directly using MTMC trackers to profile spatio-temporal correlations in the history video is computationally expensive, neutralizing the savings from the search pruning. 
This is because unlike single-target tracking, a MTMC tracker will track \textit{all entities} in the dataset. 
To limit the profiling overheads, we explore the trade-off between the robustness of offline profiling and the accuracy of subsequent single-target cross-camera tracking using the generated model.
In particular, the profiling cost can be reduced by labeling fewer frames with the MTMC tracker (\eg by selecting a lower frame sampling rate or choosing a smaller subset of the data to label).
At first glance, this will likely reduce the search accuracy as the spatio-temporal correlations is based on a sampled subset of entities.
In practice, however, we found that despite labeling fewer frames for the profiling, our precision and recall drops are only mild, and thus our solution of labeling fewer frames significantly reduces the profiling cost {\em without} impacting accuracy.
We empirically show this in \S\ref{sec:profile-eval}.

Finally, ReXCam needs to cope with potential changes in the spatio-temporal correlations (\eg a road work may block a busy segment, which can reduce the correlation between two cameras).
These `changes are relatively infrequent, but when they do happen, ReXCam can automatically detect them and initiate re-profiling.
In particular, ReXCam tracks the number of objects that are missed in the normal pruned search but detected in the subsequent replay search (in an ``uncorrelated'' time interval or camera), and triggers a re-profiling of the spatio-temporal correlations between the {\em corresponding cameras} when there is a spike in pruning errors.
Note that the error in the spatio-temporal profile {\em during the} re-profiling will not affect \name's inference, but only increase latency because the replay search handles the errors.

\section{System Implementation \& Deployment}
\label{sec:implementation}

\begin{figure} [t]
\includegraphics[width=0.45\textwidth]{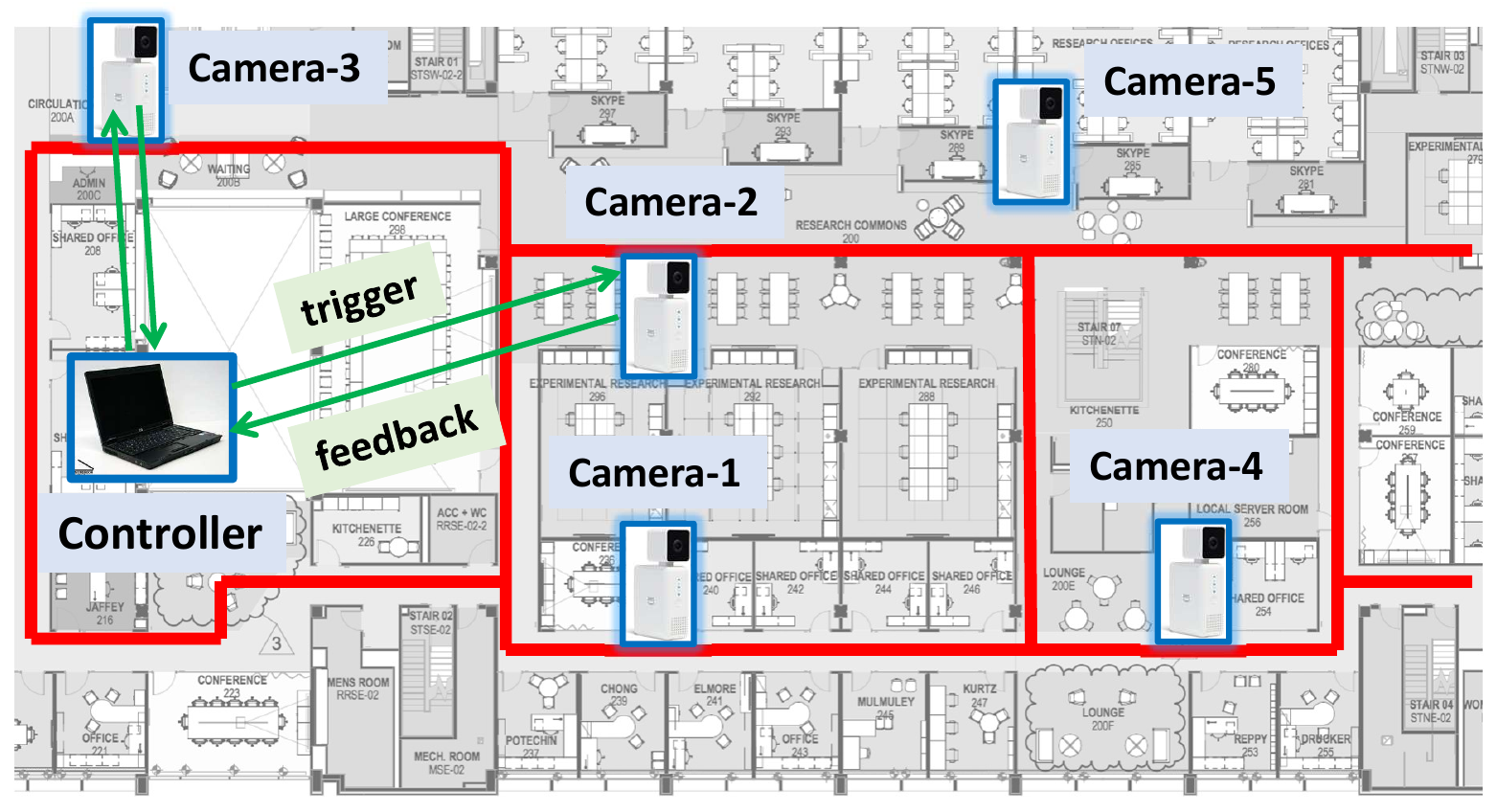}
	\vspace{-4mm}
	\caption{\small \bf \name testbed deployment at AnonCampus with five AWS DeepLens smart cameras. The red lines show the walkways in the building, and we learn the spatio-temporal correlation of people traversing the walkways. The {\em controller} and all the cameras exchange ``trigger'' and ``feedback'' messages.}
	\label{fig:testbed}
	\vspace{-2mm}
\end{figure}

We implement \name with 1.5K line of Python code over AWS DeepLens smart cameras~\cite{awsdeeplens}. 
Each DeepLens camera runs Ubuntu OS-16.04 LTS, and is equipped with an Intel Gen9 GPU and Intel Atom Processor CPU, 8{GB} RAM, and 16{GB} built-in storage. Our testbed includes five such cameras connected to each other via Wi-Fi and deployed on AnonCampus (Figure \ref{fig:testbed}). In our testbed, video analytics modules (object detection, re-id) run on DeepLens's on-chip GPU and CPU. The testbed of smart cameras contrasts the alternate model for video analytics using nearby edge boxes (\S\ref{subsec:setup}). 

We use a laptop (connected to the same Wi-Fi network as the cameras) to run the {\name} {\em controller}. 
The \name controller is responsible for profiling (\S\ref{sec:training}) and maintaining the spatio-temporal model of correlations among cameras. The connectivity between the controller and the cameras is only to exchange ``control messages'' and not video data. We implement two main control inferences (Figure \ref{fig:testbed}): 
\begin{enumerate}\vspace{-.07in}
\item A {\em trigger} message from the controller to a camera triggers the camera to start (or stop) searching for a specified query identity in its video within a specified time interval. The trigger message can also be used to initiate search in history videos for replay search (\S\ref{sec:replay}). \vspace{-.07in}
\item A {\em feedback} message from a camera to the controller notifies the controller on an interesting incident (\eg the specified identity has just been detected, or left the camera's view) in real-time. A feedback follows an activation message and is sent as soon as the incident occurs.
\vspace{-.05in}
\end{enumerate}

\noindent{\bf Fault tolerance:} The cameras broadcast a heartbeat every few seconds to the controller to handle instances of cameras failing. The \name controller can be replicated for resilience.
The only persistent state held by the \name controller is the model of spatio-temporal correlations, which is backed up, and is updated only at coarse timescales. 
The spatio-temporal pruning algorithm (Algorithm~\ref{alg:tracking-st}) is also stateless, and triggered by feedback messages from the cameras. 



\section{Evaluation}
\label{sec:evaluation}

Our evaluation of \name shows the following highlights. 

1) \name's compute savings on the 8-camera DukeMTMC dataset is $8.3\times$ (which is $\sim90\%$ of the potential; \S\ref{sec:potential}). \name also improves precision from 51\% to 90\%.
On the larger simulated dataset of 130 cameras from Porto, our savings grow with the number of cameras. (\S\ref{sec:results}, \S\ref{sec:eval-frs})

2) Deployment on the 5-camera testbed with AWS DeepLens cameras leads to $3.4\times$ savings in compute. (\S\ref{sec:results})

3) \name's optimizes to keep the profiling costs small without impacting the precision and recall. (\S\ref{sec:profile-eval})

We evaluate \name for single-camera analytics in \S\ref{sec:iddet-eval}.

\subsection{Methodology}
\label{subsec:methodology}
   
\ \ \ {\bf A. Datasets ---} 
We evaluate ReXCam on three datasets.

\noindent{\em 1) AnonCampus dataset} (\S\ref{sec:implementation}) consists of 35 minutes of $1080$p video recorded at 24 frames per second, captured by five DeepLens cameras deployed in a school building (see {Figure \ref{fig:testbed}}). The dataset is manually labeled with person identities.

\noindent{\em 2) DukeMTMC dataset} is a video surveillance dataset with footage from eight cameras installed on the Duke University campus (see {Figure \ref{fig:duke-map}}). The data consists of 85 minutes of $1080$p video from each camera recorded at 60 frames per second. In all, the footage contains over 2,700 unique identities and over 4 million person detections (all labeled).

\begin{figure} [t!]
	\includegraphics[width=0.48\textwidth]{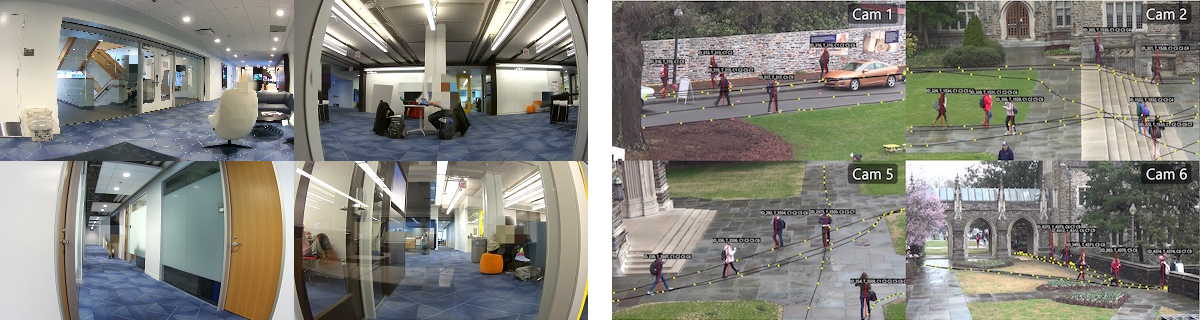}
	\vspace{-4mm}
	\caption{\small \bf Example snapshots from AnonCampus (left) and DukeMTMC\cite{DukeMTMC} (right) cameras. \protect}
	\label{fig:dataset}
	\vspace{-2mm}
\end{figure}

{Figure~\ref{fig:dataset}} shows snapshots from eight different cameras (four each) from the AnonCampus and DukeMTMC datasets.

\noindent{\em 3) Porto dataset} is generated from 1,710,671 trajectories obtained from 442 taxis running in the city of Porto, Portugal between Jan. 2013 and June 2014 \cite{PortoDataset}. Each trajectory contains timestamps and GPS coordinates sampled every 15 seconds. To emulate cross-camera tracking, we {\em manually pin} 130 cameras at intersections of the city (we get the cameras' coordinates from Google Maps) and set each camera's field-of-view to be a square area centered at the camera with length $l= 100$m. We assume the accuracy of object detection and re-id equal to the values reported in DukeMTMC-reID~\cite{DukeReID} for objects in the camera's view. The main objective is to measure \name's gains in a large city-wide setting of cameras.

{\bf B. Models ---} 
For our re-id model, we use an open-source, ResNet-50-based implementation of person re-id \cite{DPR}, trained in PyTorch on a subset of the Duke dataset called DukeMTMC-reID \cite{DukeReID}. We then implement our tracking (Algorithms \ref{alg:tracking-st}), which applies this model iteratively at inference time to discover all instances of a query identity in the Duke dataset. 
Since DeepLens uses the {\sf clDNN} and Intel GPUs, we leverage {\sf person-reidentification-retail-0076} from the OpenVINO model zoo~\cite{openvino} for re-id in the AnonCampus dataset. 

To build our spatio-temporal model on unlabeled video data (simulating real deployment conditions), we apply an offline multi-target multi-camera (MTMC) tracker \cite{DeepCC} (\S\ref{sec:training}) to label every person detection in a subset of the dataset (\ie \texttt{profile} set with $16352$ frames). 
We implement a \textit{profiler} to extract spatial and temporal correlations from these labels. 

{\bf C. Workload ---} 
We run a set of 100 tracking queries, $\{q_i\}$, drawn from the \texttt{test} query partition of the DukeMTMC-reID dataset \cite{DukeReID} (20 from the AnonCampus dataset, and 100 from the Porto dataset). Each tracking query consists of multiple \textit{iterations}. Each iteration involves searching for the next \textit{instance}, $q_i^j$, of the query identity in the dataset, starting with the initial instance $q_i^0$. A tracking query terminates when no more instances can be found. Experiments on the DukeMTMC dataset were conducted on AWS EC2 {\sf p2.xlarge} instances (contains one Nvidia Tesla K80 GPU).

{\bf D. Metrics ---} We report the following four metrics which are computed over the entire query set. 
	(i) \emph{Compute cost} -- Number of video frames processed, aggregated over all queries $\{q_i\}$. 
	(ii) \emph{Recall (\%)} -- Ratio of query instances retrieved to all query instances in dataset, $q_i^j$. 
	(iii) \emph{Precision (\%)} -- Ratio of query instances retrieved to all retrieved instances, $r_i^j$. 
	(iv) \emph{Delay (sec.)} -- Lag between position of tracker and current video frame, in seconds, at the end of a tracking query. This will be 0 for a query if no replay search was performed. 

\noindent{Compute cost, recall, and precision are reported in aggregation. Delay is reported as an average value per query.}

{\bf E. Compared Schemes ---} To evaluate our spatio-temporal filtering, we compare against two schemes:


\noindent{\bf 1) Baseline (all)} - Searches for query identity $q$ in all the cameras at every frame step. Uses state-of-the-art re-id model \cite{DPR}. \textit{no} spatio-temporal filtering is utilized. 

\noindent{\bf 2) Baseline (GP)} - Searches for query identity $q$ only in the cameras that are in geographical proximity to the query camera at every frame step. Uses state-of-the-art re-id model \cite{DPR}. For DukeMTMC dataset, we manually set pairs of neighboring cameras using Figure~\ref{fig:duke-map} while for Porto dataset, we set geographical proximity threshold to $4l$ (where $l=100$m).  

\noindent{\bf 3) ReXCam} - Searches for query identity $q$ only on cameras that are currently \textit{spatio-temporally correlated} with $c_q$ (as per Algorithm \ref{alg:tracking-st}). 
The same person re-id model is used as in the baseline \cite{DPR}. 
We consider various versions of Equation~\ref{eq:model}, corresponding to different spatio-temporal filters. Each version is coded as S$s$-T$t$, where $s$ indicates the spatial filtering threshold and $t$ indicates the temporal filtering threshold. Higher values of $s$ and $t$ indicate more aggressive filtering (no $t$ value indicates no temporal filtering and helps measure the gains of spatial filtering alone). For instance, \textbf{S5-T2} filters cameras that receive \textbf{$<$5\%} of the traffic from query camera $c_q$. In addition, its filter frames outside the time window containing the \textbf{first 98\%} of traffic from $c_q$.

\subsection{Spatio-temporal filtering gains}
\label{sec:results}

\begin{figure}[t!]
	\includegraphics[width=0.48\textwidth]{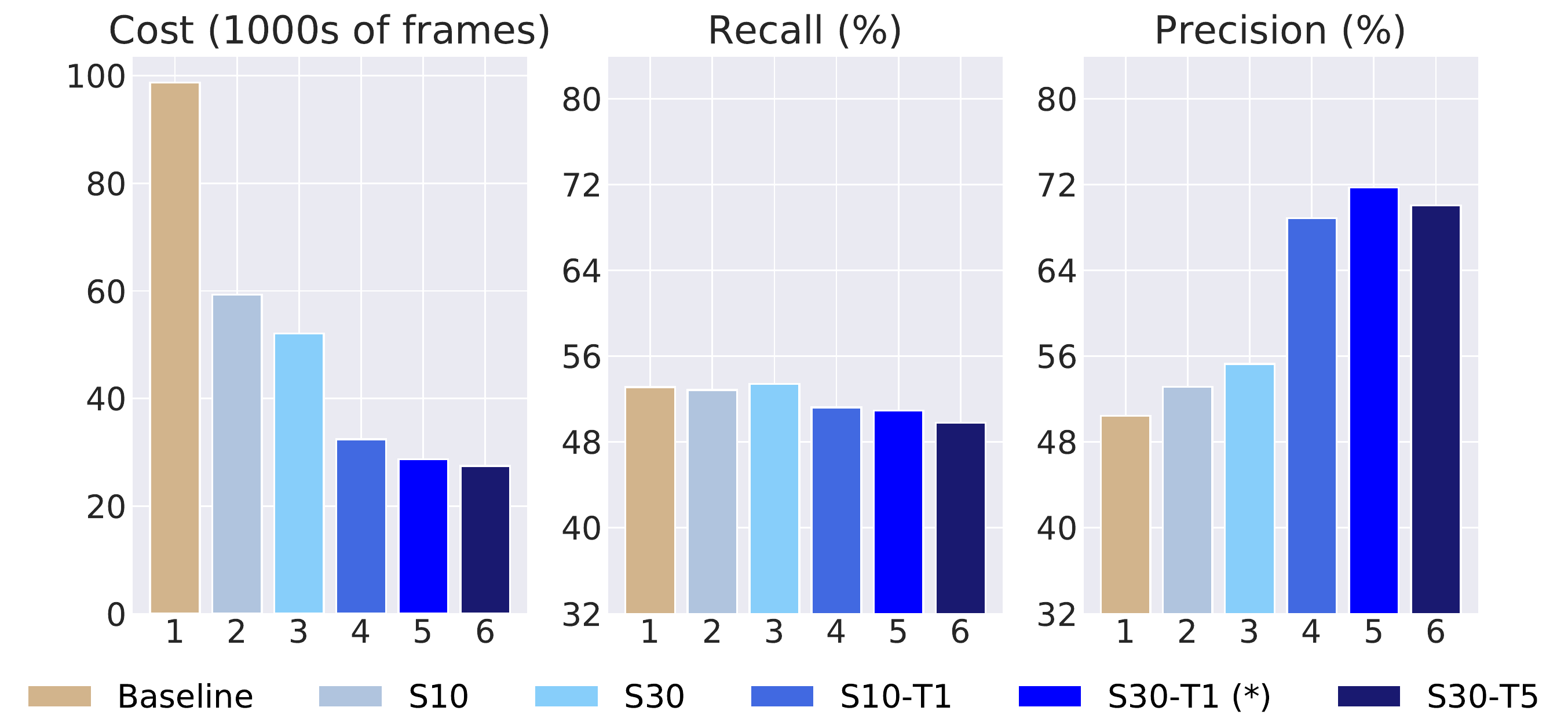}
	\vspace{-6mm}
	\caption{\small \bf Results for all-camera baseline (tan) vs. five versions of ReXCam (blues) on the AnonCampus dataset. We argue S30-T1 (*) offers the best trade-off on all metrics.}
	\label{fig:st-results-chicago}
	\vspace{-0mm}
\end{figure}

\begin{figure}[t!]
	\includegraphics[width=0.48\textwidth]{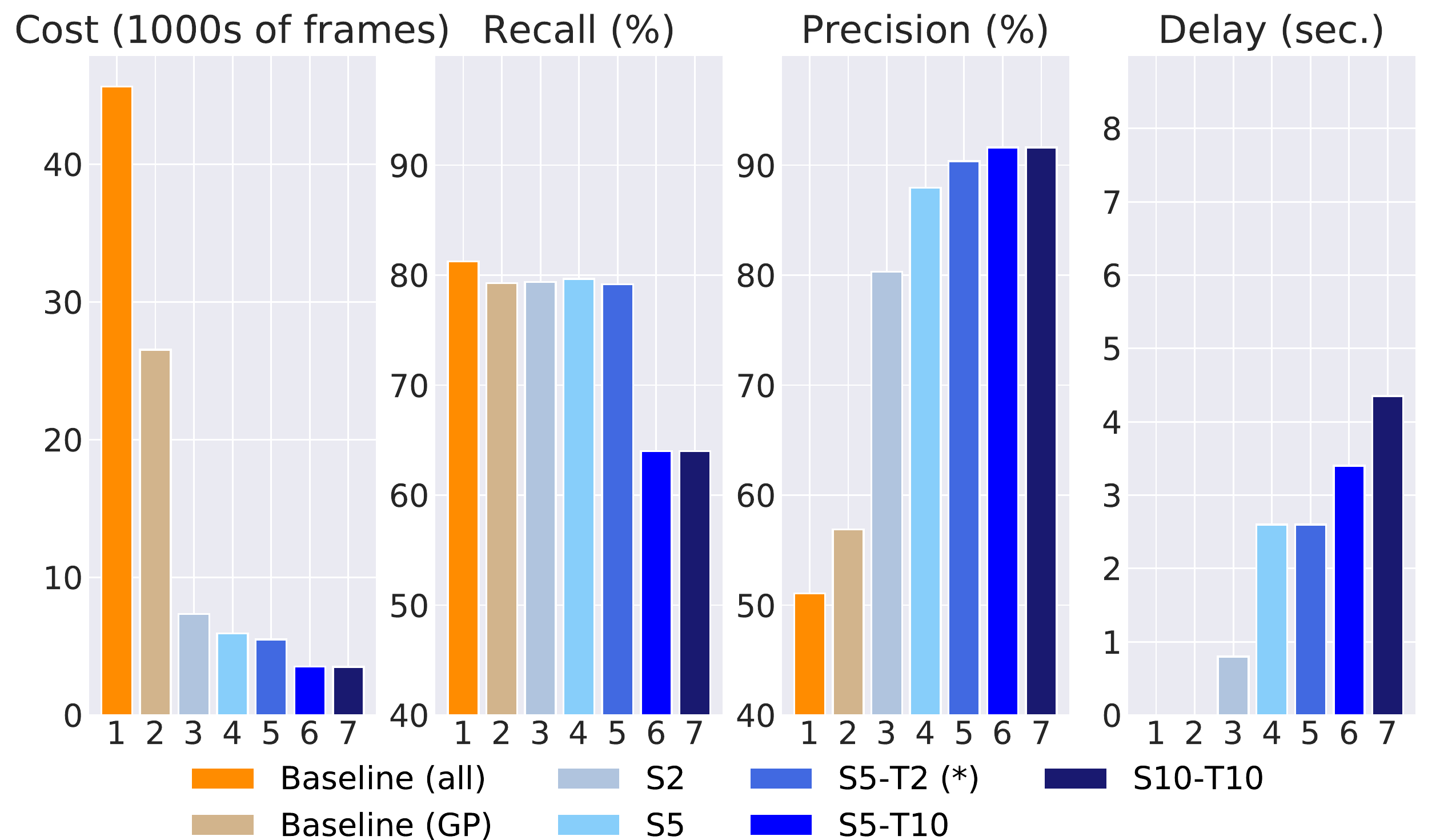}
	\vspace{-6mm}
	\caption{\small \bf Results for all-camera baseline (orange), geo-proximity baseline (tan) vs. five versions of ReXCam (blues) on the DukeMTMC dataset. We argue S5-T2 (*) offers the best trade-off on all metrics.}
	\label{fig:st-results-duke}
	\vspace{-2mm}
\end{figure}

\begin{figure}[t!]
	\includegraphics[width=0.48\textwidth]{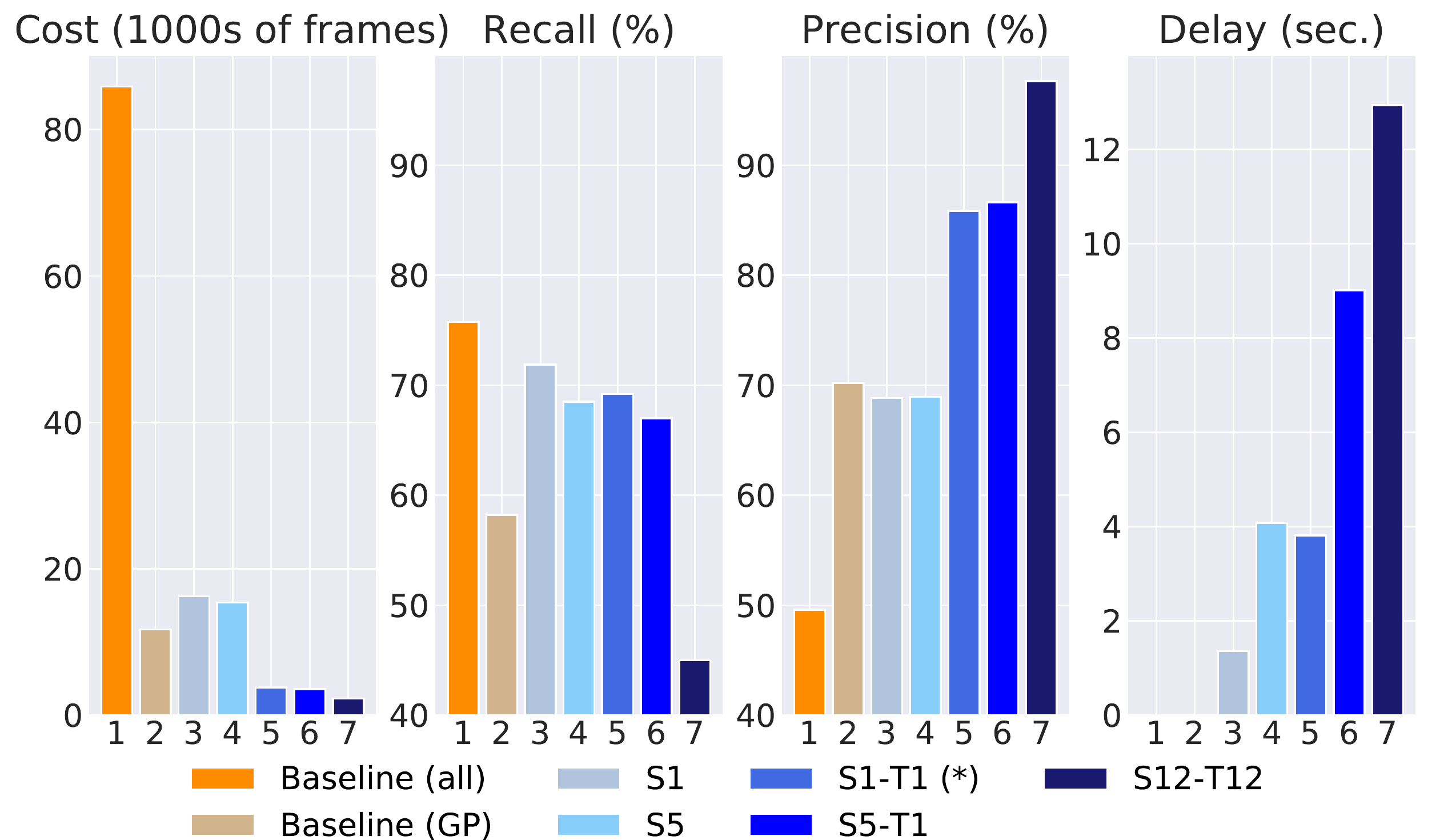}
	\vspace{-6mm}
	\caption{\small \bf Results for all-camera baseline (orange), geo-proximity baseline (tan) vs. four versions of ReXCam (blues) on the Porto dataset (130 cameras).}
	\label{fig:st-results-porto}
	\vspace{-4mm}
\end{figure}

{Figure \ref{fig:st-results-chicago}}, {Figure \ref{fig:st-results-duke}} and {Figure \ref{fig:st-results-porto}} compare the performance of the baseline and various ReXCam versions on three datasets, respectively. We find that ReXCam \textit{significantly outperforms both baselines}, by (1) reducing compute cost and (2) improving precision, while maintaining comparable recall. It is noteworthy that the best thresholds for \name is dependent on the dataset. ReXCam versions \textbf{S30-T1}, \textbf{S5-T2}, \textbf{S1-T1} offer the best trade-off between compute cost, recall, precision, and delay in the three datasets, and in general have to be tuned. We term these schemes \textbf{ReXCam-O}(ptimal).


	\noindent\textbf{1) Compute cost} -- Baseline (all) is by far the most compute-intensive, processing 98,760 frames for 20 queries and 45,638/85,890 frames for 100 queries on the DukeMTMC/Porto dataset, respectively. Baseline (GP) saves the cost quite a bit but its performance fluctuates on different settings due to the discrepancy between spatial correlation and geographical proximity (as also pointed out in \S\ref{subsubsec:spatial}). Each successive version of ReXCam achieves lower compute cost than its predecessor. For instance, in {Figure \ref{fig:st-results-duke}}, the most aggressive version of ReXCam, S10-T10, processes only 3,513 frames, and \textit{achieves 13$\times$ lower compute cost} on 8 cameras than the all-camera baseline. Similarly, a maximal value of \textit{3.6$\times$ compute savings} can be achieved in {Figure \ref{fig:st-results-chicago}}. 

	In comparison, \textbf{ReXCam-O} processes 28,680/5,500/3,776 frames, which translates to $3.4\times$/$8.3\times$/$23\times$ lower cost than the all-camera baseline in the five-camera (AnonCampus), eight-camera (DukeMTMC), and 130-camera (Porto) dataset.

	\noindent\textbf{2) Recall (\%)} -- Compared with both baselines, recall of the ReXCam versions \textit{declines} slightly when spatial/temporal filtering is introduced. 
	In {Figure \ref{fig:st-results-duke}}, for example, baseline (all) achieves recall of 81.3\%. Both spatial-only schemes achieve 79.3\% recall. \textbf{ReXCam-O} achieves 79.7\%, a 1.6\% drop from the baseline. Similar patterns are observed in {Figure \ref{fig:st-results-chicago}} and {Figure \ref{fig:st-results-porto}}. 
	The reason why recall becomes lower in the AnonCampus deployment is because of the increased instances of occlusions in indoor environments (see {Figure \ref{fig:dataset}}). Note that in {Figure \ref{fig:st-results-porto}}, recall drops significantly from baseline (all) to baseline (GP), as a number of relevant cameras are mistakenly excluded by geographical proximity-based pruning. 


	\noindent\textbf{3) Precision (\%)} -- Baseline (all) achieves precision of 50.4\%, 51.1\% and 49.6\% on three datasets, respectively. All versions of ReXCam improve on this, but \textbf{ReXCam-O} in particular achieves 71.7\%/90.4\%/85.8\% precision, which is a \textit{gain of 21.3\%/39.3\%/36.2\% over the baseline}. Compared with baseline (GP), precision gain from \textbf{ReXCam-O} remains as high as \textit{33.5\%/15.6\%} on the DukeMTMC and Porto dataset. 
	Higher precision is a key benefit of spatio-temporal filtering for cross-camera video analytics. By searching fewer irrelevant cameras, and fewer irrelevant frames, ReXCam is less likely to declare matches that do not actually match the query. 


	\noindent\textbf{4) Delay (sec.)} -- Here we report total cumulative lag (lag in the absence of replay search (\S\ref{sec:replay})), averaged over all queries. We do not report the delay from the AnonCampus deployment since among all 20 queries, only one needed replay search. 
	For both DukeMTMC and Porto results, we find that delay increases with more spatial or temporal pruning. This is expected as there are more instances of misses. \textbf{ReXCam-O}, in particular, incurs moderate delay -- less delay than S5-T1 and S5-T10 but more delay than spatial-only filtering.

Given this analysis, \textbf{ReXCam-O} offers a favorable tradeoff between the four metrics -- achieving nearly the lowest compute cost ($3.4\times$/$8.3\times$/$23\times$ lower), nearly the highest precision ($21.3\%$/$39.3\%$/$36.2\%$ higher), competitive recall ($2.2\%$/$1.6\%$/$6.5\%$ lower), and moderate lag ($\approx$3.2s), when compared to the locality-agnostic, all-camera baseline. 
Next, we analyze the impact of two key factors on \name.

\begin{figure}[t!]
	\includegraphics[width=0.48\textwidth]{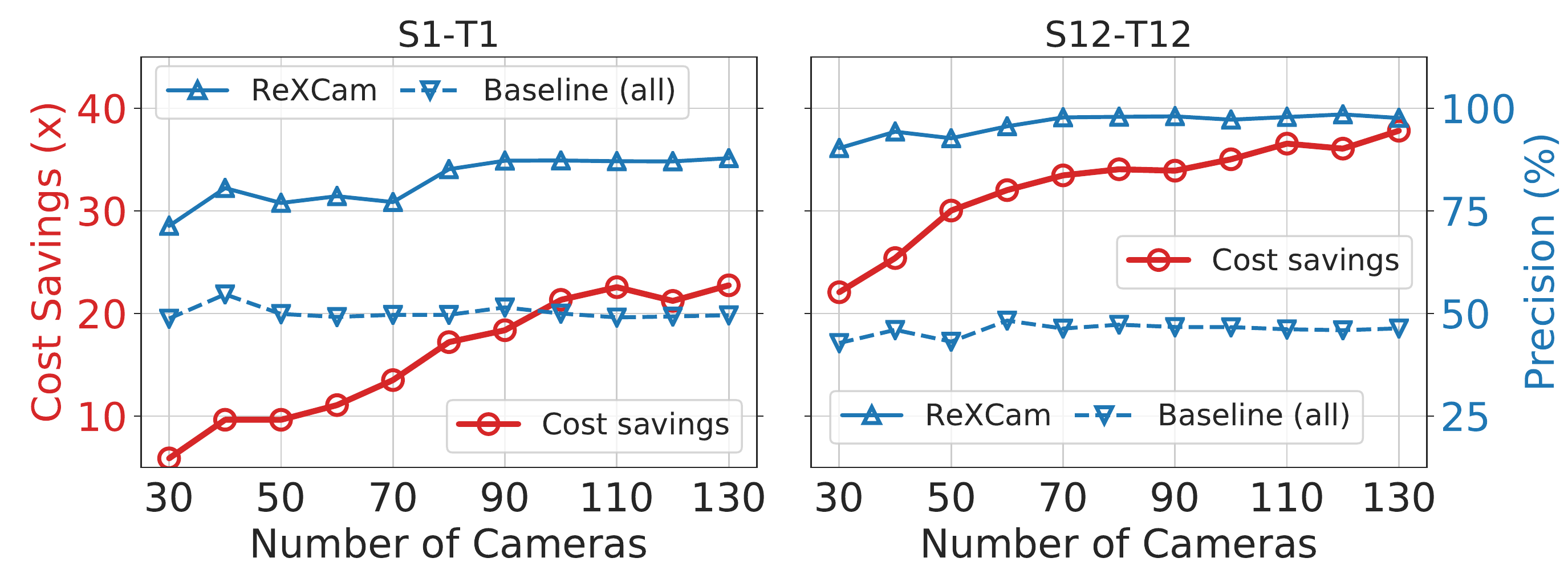}
	\vspace{-6mm}
	\caption{\small \bf Cost savings vs. number of cameras (Porto dataset).}
	\label{fig:cam-num-results}
\end{figure}

\textbf{Large-scale camera data:} The key objective of using the trajectories from the Porto dataset was to experiment on \name's gains at scale (\S\ref{subsec:methodology}); unfortunately there are no video datasets available for hundreds of cameras. Figure~\ref{fig:cam-num-results} shows cost savings and precision of ReXCam/Baseline (all) with increasing number of cameras. Cost savings steadily grows with increasing number of cameras, achieving up to $38\times$ lower cost than baseline (all) in ReXCam S12-T12 for 130 cameras. We believe this is an encouraging result for \name's value for large camera deployments. All through, ReXCam maintains a $34.5\%$ gain on precision with little impact on recall. 
\begin{figure}[t!]
	\includegraphics[width=0.5\textwidth]{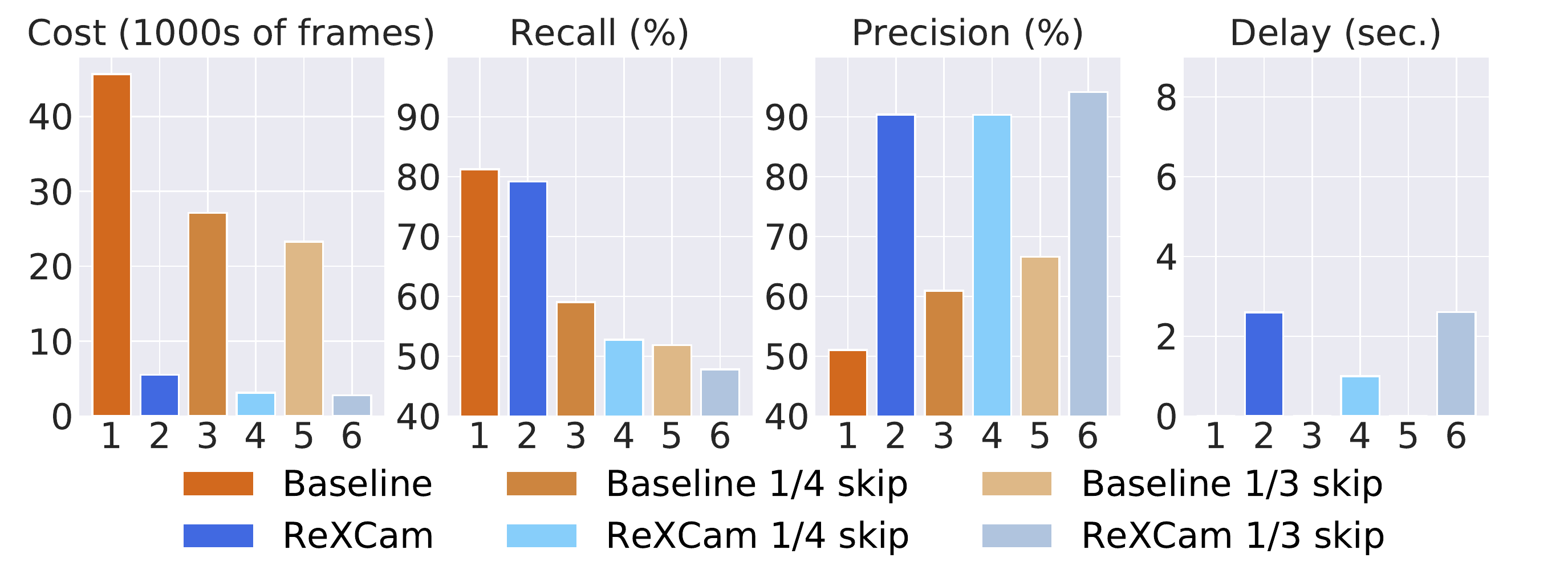}
	\vspace{-6mm}
	\caption{\small \bf Results for all-camera baseline vs. ReXCam S5-T2 on the DukeMTMC dataset with frame skipping.}
	\label{fig:st-results-duke-skip}
 	\vspace{-2mm}
\end{figure}

\textbf{Frame skipping:} Frame sampling is a key technique in prior work \cite{VideoStorm, Chameleon, Focus} to make {\em single-camera} analytics cheaper. Such techniques are orthogonal to \name's spatio-temporal pruning for cross-camera analytics, and we quantify our point. Figure~\ref{fig:st-results-duke-skip} measures the impact of frame skipping---uniformly skip one in 3 frames, and one in 4 frames---on both baseline (all) and ReXCam. As shown in the figure, ReXCam maintains a much lower compute cost in both skipping cases. Specifically, the cost savings are $8.6\times$ and $8.4\times$, which is in the same ballpark as without frame skipping of $8.3\times$, thus showing the orthogonality of frame skipping to \name.

\subsection{Replay search}
\label{sec:eval-frs}
In this section, we evaluate the effectiveness in reducing lag in replay search using the two proposed schemes from \S\ref{sec:replay}:

	\noindent\textit{Skip frame mode} - Employ a $\frac{x}{2}$ frame sampling rate to increase throughput on historical frames, at the price of lower accuracy (via missed detections). (\textbf{2x skip})
	
	\noindent\textit{Parallelism mode} - Employ a $2x$ frame processing rate to increase throughput, at the price of increased compute cost (via increased resource usage). (\textbf{2x ff})

Both schemes are applied to \textbf{ReXCam-O}, and compared to (a) the all-camera baseline and (b) \textbf{ReXCam-O} with the default \textit{real-time} replay search, which incurs 2.6s of delay.
\begin{figure}
	\includegraphics[width=0.48\textwidth]{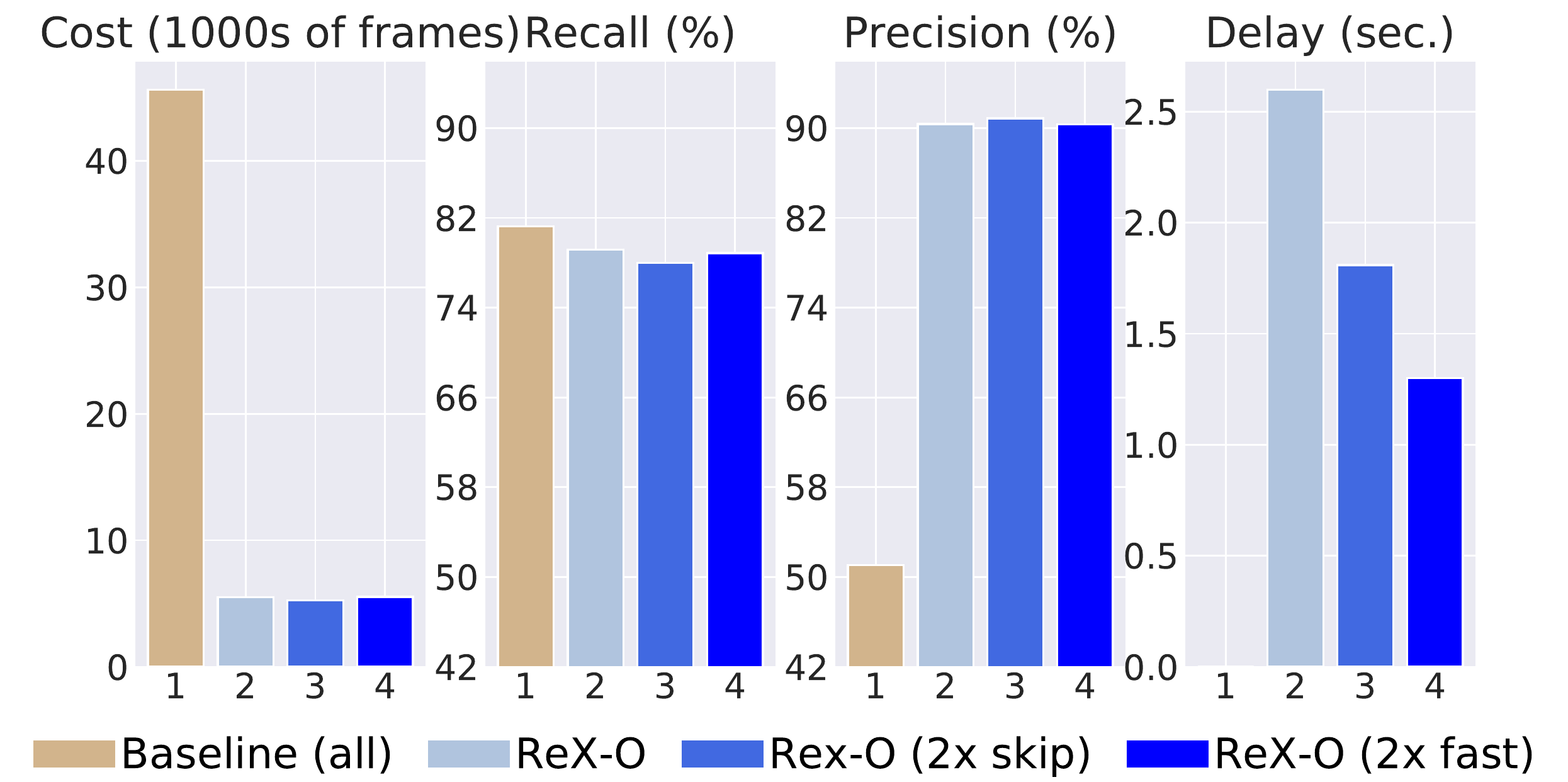}
	\vspace{-6mm}
	\caption{\small \bf Replay search. Schemes compared: baseline, ReXCam-O (normal replay search), ReXCam-O ($2\times$ skip), ReXCam-O ($2\times$ fast-forward). Scheme $2\times$ skip outperforms $2\times$ fast-forward on both compute cost and delay.}
	\label{fig:frs-results}
	\vspace{-2mm}
\end{figure}

As {Figure \ref{fig:frs-results}} shows, both \textbf{2x skip} and \textbf{2x ff} achieve delay reductions, decreasing final cumulative lag to $1.8$s and $1.3$s, respectively. The reason why \textbf{2x skip} doesn't halve the delay is due to the skipped query instances during the first round of replay search where s$_\text{thresh}$ and t$_\text{thresh}$ decreased by a factor of 10. Also, delay reductions from \textbf{2x skip} and \textbf{2x ff} come with different tradeoffs. \textbf{2x skip} reduces recall by 1.2\% to 78.0\%, but \textit{increases precision} from 90.37\% to 90.87\% and \textit{increase compute cost savings} from $8.30\times$ to $8.68\times$ better than the baseline (by processing fewer historical frames). \textbf{2x ff} does not impact recall and precision, but reduces compute cost savings from $8.30\times$ to only $8.27\times$ better than the baseline.


\subsection{Profiling cost vs. tracking accuracy}
\label{sec:profile-eval}
Profiling cost increases with the number of frames that must be processed by the MTMC tracker (\S\ref{sec:training}). We investigate the trade-off between profiling cost and subsequent tracking accuracy. Specifically, we test whether we can build a precise spatio-temporal model on smaller subsets of the training data obtained by uniformly sampling the frames. We apply a sampling rate of $8\times$, $6\times$, $4\times$, $2\times$, and $1\times$ (using $X$ in $8$ frames) in the \texttt{profile} partition of the Duke dataset (\S\ref{subsec:methodology}) for profiling, which translates to correspondingly lower profiling costs.
\begin{figure}[t!]
	\includegraphics[width=0.47\textwidth]{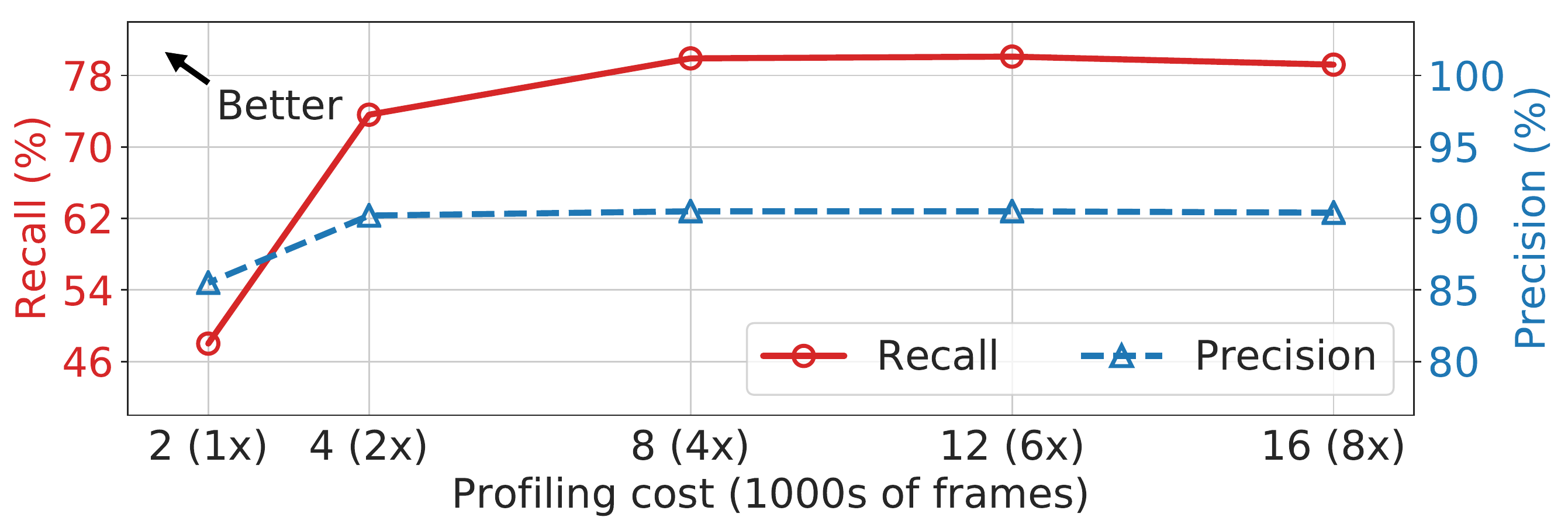}
	\vspace{-3mm}
	\caption{\small \bf Offline profiling cost vs. online recall. Profile intervals compared (in minutes of data used per camera): $49.4$ min. (full), $37.1$ min., $24.7$ min. (half), $12.4$ min., $6.2$ min.}
	\label{fig:profile-results}
	\vspace{-2mm}
\end{figure}

As {Figure \ref{fig:profile-results}} shows, recall of \name\ during live tracking reaches the maximum of 80.1\% with $6\times$ sampling, \ie when half of the frames are labeled for offline profiling to obtain the spatio-temporal model. Interestingly, on either side of this, the recall falls. On the left side, the drop is caused by insufficient amount of profiling data. On the right side, the small drop is because extra data results in a spatial-temporal model being overfit to the \texttt{profile} partition. This experiment indicates that spatial-temporal model can be built on a reasonably small set of training data (\ie 37.1 min). However, the exact amount of data to train the spatial-temporal model varies among datasets, and thus should be chosen carefully. Precision remains stable ($\sim$90\%) in {Figure \ref{fig:profile-results}} when more than 4K (\ie $2\times$ sampling) frames are used for training.

If we combine the profiling cost with the cost of the live video analytics, we see that ReXCam would need to run {\em only} 34 live tracking queries to break-even with locality-agnostic tracking (calculations omitted). This represents a small fraction of the expected annual workload in large video analytics operations \cite{Vigil, VideoStorm} that track many hundreds of thousands of queries. Hence \name's profiling costs are small and will not dent the gains, leaving it to remain sizable.


\subsection{Identity detection}
\label{sec:iddet-eval}

\begin{figure}[t!]
	\includegraphics[width=0.48\textwidth]{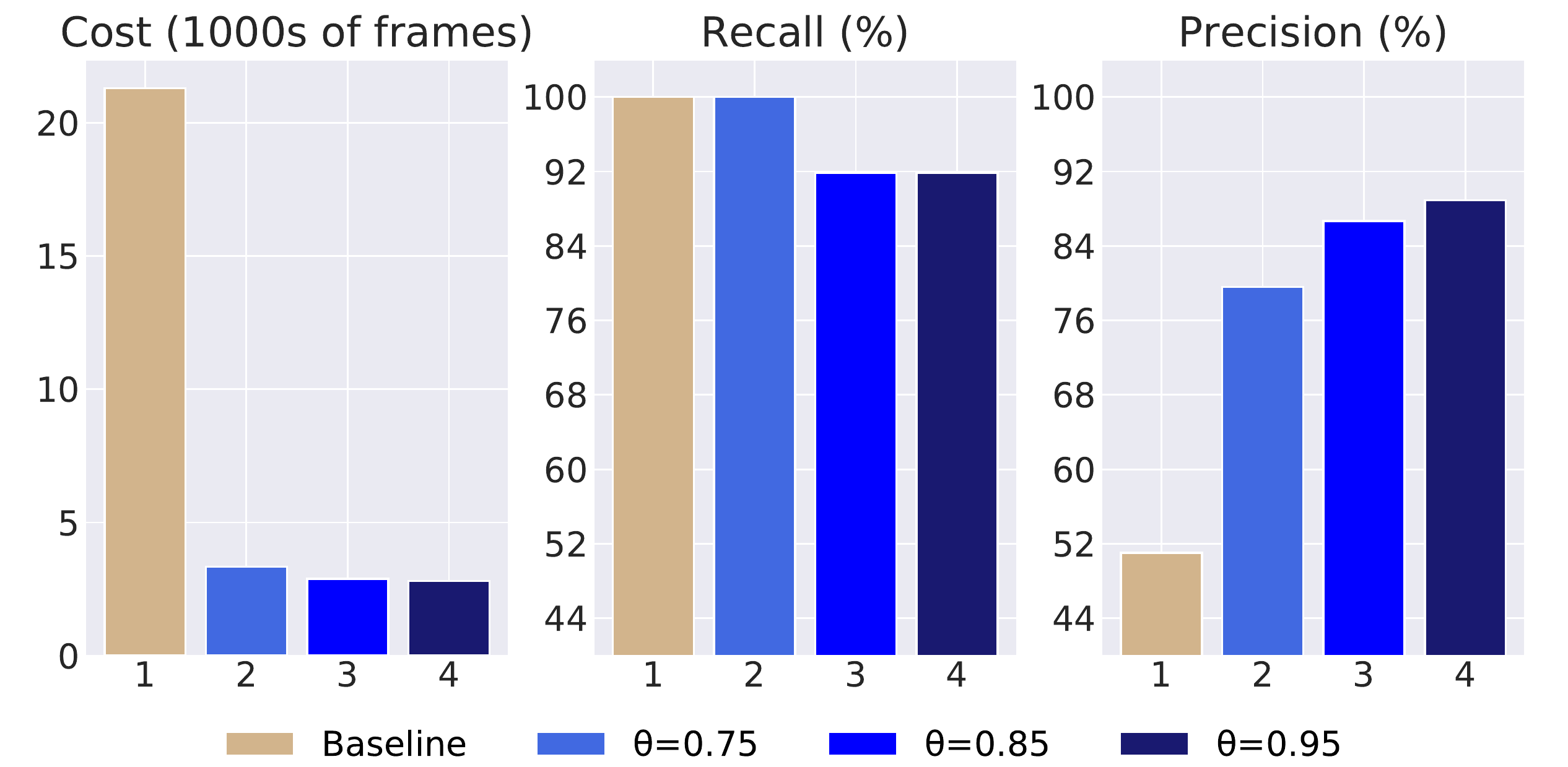}
	\vspace{-6mm}
	\caption{\small \bf Identity detection. All-camera baseline (tan) vs. three versions of ReXCam (blues) on the DukeMTMC dataset.}
	\label{fig:st-results-iddetection}
	\vspace{-4mm}
\end{figure}

Lastly, we evaluate \name's spatio-temporal pruning on identity detection, the {\em single-camera} application described in \S\ref{subsec:detection}. As Figure~\ref{fig:st-results-iddetection} shows, \name\ achieves as high as $7.6\times$ cost reduction with $\theta=0.95$ on the 8-camera DukeMTMC dataset ($\theta$ is the likelihood threshold for searching a camera's stream). Similar to trends in cross-camera tracking, the gain on precision far outweighs the drop on recall. In fact, for $\theta=0.75$, recall does not drop at all while precision improves by $28\%$ even as cost savings stay at $6.6\times$. This experiment shows the generality of applying \name for both cross-camera as well as single-camera applications. 

\section{Related Work}

\noindent{\bf Video Analytics Systems.} A sizable body of work on video analytics has emerged recently \cite{Optasia, VideoStorm, NoScope, Focus}. 
Chameleon exploits correlations in camera content (\eg velocity of objects) to amortize {\em profiling costs}, but not the cost of the video analytics itself \cite{Chameleon}. These works leave three problems unexplored, each of which ReXCam addresses.
First, they focus on {\em single-frame} tasks (\eg object detection and classification), which are stateless. In contrast, surveillance applications, like the real-time tracking we focus on, involve multi-frame tracking, where future questions depend on past inference results. 
Second, they study {\em single camera} analytics. 
Thus, they do not explore the complexities involved in cross-camera {\em inference} on live video (\eg occlusions) that define applications such as person re-id. 
Third, in contrast to classification tasks, many security applications search for new object instances (\eg a suspicious person) where the training data is skewed toward negative examples. 
Our use of correlations, \ie movement across cameras, however, yields substantial accuracy gains.

\noindent{\bf Efficient Machine Learning.} 
Improving ML models using model compression \cite{DeepCompress, PruningFilters}, compact architectures \cite{SqueezeNet, NIN}, knowledge distillation \cite{Distill, DoDeepCNNs, OnlineDistill}, and model specialization \cite{NoScope, Focus} is orthogonal to \name, which would gain from any efficiency improvement of the models (\eg for re-id).

Unlike systems that tradeoff model resources and accuracy \cite{MCDNN, NestDNN, Clipper, Vigil, Glimpse, DeepMon, DeepEye, Starfish, Mainstream}, ReXCam entails a new approach: instead of running cheaper models, we run inference on \textit{less data} by using spatio-temporal correlations. 

\noindent{\bf Computer Vision.} Techniques for person re-id and multi-target, multi-camera (MTMC) tracking make the following contributions: (1) new datasets \cite{DukeMTMC, PRW, PersonSearch, MSMT17}, (2) new neural network architectures \cite{PRW, PersonSearch, MSMT17, ksray2017}, or (3) new training schemes \cite{zhu2018, DukeFeatures, PRW, PersonSearch}.
However, past computer vision work do not address the {\em inference cost} of re-id and MTMC tracking \cite{Geo1, Geo2, Geo3, Geo4, Geo5}, nor does it study \textit{online} tracking (iterated re-id), a key application of interest in camera systems. 

\noindent{\bf Visual Data Management.} Image and video databases explore the use of classical computer vision techniques to index video efficiently \cite{Chabot, QBIC, VDBMS, Jacob, BrowsingDB}. \textit{Cross-camera inference} with CNNs on \textit{live video} 
entails substantially different challenges than the target domain of these works.

\noindent{\bf Mobility Modeling.} Mobility modeling and prediction has long been a topic of interest in mobile computing. Studies have shown promising results in generating human/vehicle mobility models from call detail records~\cite{isaacman12mobisys,zhang14mobicom}, wireless signals~\cite{yoon06mobisys}, social media~\cite{10.1371/journal.pone.0131469}, and transactions in transportation systems~\cite{yang16mobisys,zhang14mobicom}. While none of these works apply mobility models to video analytics, \name could benefit from their techniques on building accurate spatial-temporal models. 


\section{Conclusions}

Cross-camera analytics is a computationally expensive functionality that underpins a range of real-world video analytics applications, from suspect tracking to intelligent retail stores. We presented \name, a system that leverages a \textit{learned model of cross-camera correlations} to drastically reduce the size of the inference time search space, thus reducing the cost of cross-camera video analytics. 
\name directs its search towards the camera streams that likely contain the identity being tracked, while gracefully recovering from (rare) misses using a replay search on historical videos. 
Our results are promising: \name reduces compute workload by $8.3\times$ on the 8-camera DukeMTMC dataset, and improve inference precision by 39\%. On a simulated dataset of 130 cameras, its gains grow with the number of cameras. We have deployed a five camera testbed on campus, which we plan to expand for further experiments.
\vspace{+4mm}

\bibliographystyle{plain}
\bibliography{reference}

\end{document}